\documentclass[preprint,12pt]{elsarticle}

\usepackage{wrapfig,enumerate}

\usepackage{epsfig,subfigure}
\usepackage{wrapfig}

\usepackage[cmex10]{amsmath}
\usepackage{amssymb}

\def\PRSzero    {\mbox{\rm PRS0}}
\def\PRSone    {\mbox{\rm PRS1}}
\def\RIFS    {\mbox{\rm RIFS}}
\def\CIFS    {\mbox{\rm CIFS}}

\begin{document}

\begin{frontmatter}

\title{On Efficiency and Validity of Previous Homeplug MAC Performance Analysis}

\author{C. Cano}
\ead{cristina.cano@nuim.ie}
\author{D. Malone}
\ead{david.malone@nuim.ie}

\address{Hamilton Institute \\ National University of Ireland, Maynooth \\ Co. Kildare, Ireland.}

\begin{abstract}

The Medium Access Control protocol of Power Line Communication networks (defined in Homeplug and IEEE 1901 standards) has received relatively modest attention from the research community. As a consequence, there is only one analytic model that complies with the standardised MAC procedures and considers unsaturated conditions. We identify two important limitations of the existing analytic model: high computational expense and predicted results just prior to the predicted saturation point do not correspond to long-term network performance. In this work, we present a simplification of the previously defined analytic model of Homeplug MAC able to substantially reduce its complexity and demonstrate that the previous performance results just before predicted saturation correspond to a transitory phase. We determine that the causes of previous misprediction are common analytical assumptions and the potential occurrence of a transitory phase, that we show to be of extremely long duration under certain circumstances. We also provide techniques, both analytical and experimental, to correctly predict long-term behaviour and analyse the effect of specific Homeplug/IEEE 1901 features on the magnitude of misprediction errors.  

\end{abstract}


\end{frontmatter}

\section{Introduction}

% PLC Communications

Data transmission using electrical wires, known as Power Line Communication (PLC), has the potential to become a useful complement and strong competitor to wireless networking solutions. The appeal of PLC relies in the high data rates it can deliver, its low deployment cost (as it is easy to retrofit) and allows communication through obstacles that commonly degrades wireless signals. Additionally, it also provides a low-cost alternative to complement existing technologies to reach ubiquitous coverage. For instance, as a backhaul for wireless sensor networks or small cells.

Research efforts have been mostly focused on the physical layer as the characteristics of PLC channels (including fading, impulsive noise and hidden/exposed terminal problems) impose several challenges on physical aspects \cite{ferreira1999power}. However, they also have an impact on the Medium Access Control (MAC) protocol, which, in contrast, has not received much attention. 
 
% Homeplug MAC Overview

PLC standards (we focus on Homeplug \cite{HomeplugStd} and IEEE 1901 \cite{IEEE1901}) define a MAC procedure similar to the Distributed Coordination Function (DCF) defined in the IEEE 802.11 standard for Wireless Local Area Networks (WLANs) \cite{IEEE80211-IEEESTD1999}. PLC MAC protocols diverge from the vanilla DCF by adding a \emph{deferral counter} that reduces the attempt rate when high contention is inferred on the channel (i.e., a certain number of packets are overheard). Despite being a substantial change over DCF, this feature has not been deeply evaluated. Attempts to compare DCF and PLC MAC protocols have been made in \cite{Velloso,Li2011,vlachou2013fairness} and \cite{cano2013pimrc}. However, there is still much work to do in order to fully demonstrate the conditions under which the deferral counter improves the performance of the network.

% Homeplug Analytical Models

Contributions to the analysis of PLC MAC protocols aim to extend the research on performance evaluation of PLC networks, crucial to advance understanding and optimisation. The only relatively complete analytic model of Homeplug MAC is presented in \cite{chung2006performance}. This analytic model has been widely used \cite{yoon2008adaptive,koutny2011homeplug,pinero2011homeplug,koutny2013analysis,pinero2011realistic,yoon2013multichannel}, however we identify two significant issues. The first is the high computational complexity, making it unsuitable for online use or study of complex scenarios. Second, as we will demonstrate, the performance results obtained in the regime right before the predicted saturation point do not correspond to the long-term network behaviour. In detail, the main contributions of this work are the following:

\begin{itemize}
 \item We propose a reformulation of the analytic model for the Homeplug MAC procedure presented in \cite{chung2006performance} that facilitates a simplified method of solution. Our proposed analysis provides a 2 orders-of-magnitude improvement in runtime compared to the one presented in \cite{chung2006performance} while maintaining its accuracy. Specifically, we: \emph{i)} apply a renewal reward approach \cite{kumar05,bianchi05}, \emph{ii)} allow the most expensive operations to be precomputed and \emph{iii)} provide and evaluate the accuracy of an optional exponential approximation to the probability to defer in a given backoff stage.
 \item We demonstrate that the results right before the predicted saturation point presented in \cite{chung2006performance} correspond to a transitory phase of the system instead to the long-term behaviour. We identify the two causes of misprediction in \cite{chung2006performance}: \emph{i)} the decoupling approximation under infinite buffer size considered in the analytical model and \emph{ii)} the presence, under certain circumstances, of an extremely long (of the order of magnitude of hours) transitory phase in experimental evaluation. This is the first work that shows the long duration of that transitory phase and, indeed, the causes of its occurrence suggest it may also be present in generic random access protocols. Having identified the causes of misprediction, we provide techniques to generate valid results from analytic models which use the decoupling approximation as well as from experimental studies based on simulations.
 \item We evaluate the impact of the deferral counter on the magnitude of misprediction errors by evaluating the different solutions obtained considering: \emph{i)} the starting values of the deferral counter proposed by the standard, \emph{ii)} no deferring (as done in IEEE 802.11 DCF) and \emph{iii)} always deferring after overhearing following the proposal presented in \cite{campista2005improving}.
\end{itemize}

% Contents

The remainder of this article is organised as follows. In Section \ref{sec:background}, we review the Homeplug MAC procedure. Next, in Section \ref{sec:rel_work}, related work on PLC MAC analytic models is discussed. Our simplified analytic model is described in Section \ref{sec:model}. Then, in Section \ref{sec:demonstration}, we demonstrate that the previous performance results right before predicted saturation do not correspond to the long-term behaviour of the network. The performance evaluation, including the validation of the presented analytic model, an evaluation of the effect of the deferral counter on misprediction errors and a complexity comparison, are presented in Section \ref{sec:validation}. Finally, some final remarks are provided.

\section{Homeplug MAC Background}\label{sec:background}

% How the protocol works

The Homeplug MAC protocol is similar to the DCF defined in the IEEE 802.11 standard \cite{IEEE80211-IEEESTD1999}. Each time a node has a new packet to transmit, the backoff stage ($i \in [1,m]$)\footnote{Actually, ($i \in [0,m-1]$) but indexes have been relabelled here for clarity of illustration.} is initialised to $1$ and a random backoff is selected among $[0,W_{1}]$. The backoff countdown is frozen when activity is detected on the channel and restarted when the medium becomes idle again. The packet is actually transmitted when the backoff countdown expires. If an acknowledgement is received, the packet is considered successfully transmitted. Otherwise, the node starts the retransmission procedure: the backoff stage changes to $i=\min(i+1,m)$ and a new random backoff is selected among $[0,W_{i}]$, being $W_{i}$ the contention window of stage $i$. 

Additionally, in the Homeplug and IEEE 1901 MAC, a new counter, called the Deferral Counter (DC), is introduced. This counter is initialised at each backoff stage to $M_i$ and decreased by one after overhearing a data packet or a collision. If a new packet or a collision are overheard and the value of the DC is equal to zero, the node acts as if a collision had happened: the backoff stage is increased if it has not yet reached its maximum value and a new backoff is selected among $[0,W_{i}]$. The goal of the DC is to avoid collisions when high contention is inferred by decreasing the aggressiveness of transmission attempts.

To provide channel access differentiation, four access categories (CA) are defined CA0--3. CA3 and CA2 share $W_i$ and $M_i$ values, as do CA1 and CA0 (see Table \ref{tbl:access_categories}). Two slots (called PRS0 and PRS1) are allocated to allow nodes to announce the priority of their transmissions. The highest priority (CA3) is associated to both PRS0 and PRS1, the CA2 category is associated to PRS0 only, CA1 to PRS1 and the lowest access category (CA0) does not have any notification interval associated. Following this approach, stations know if there is a station with a frame that belongs to a higher category. In such a case, they postpone their transmissions until the high priority frames are released. However, in this work we consider the case where all stations contending for the channel belong to the same access category, as assumed in \cite{chung2006performance}.

% Graphical description - time line or Markov Chain 

\section{Related Work on Homeplug MAC Modeling}\label{sec:rel_work}

% Describe previous Analytical Model

The analytic model in \cite{chung2006performance} takes into account all the features of the Homeplug MAC procedure except the channel access prioritisation. It models the node access procedure as a 3-dimensional Markov Chain in which the backoff and deferral counters, as well as the different backoff stages are considered. To solve for the stationary probabilities of the Markov Chain an iterative numerical method is required. Then, to compute the channel access delay, the authors follow a recursive approach, i.e., the delay at every backoff stage depends on the delay of previous stages. To obtain this performance metric, further iterations are needed. Moreover, to compute the channel access probabilities as well as the delay, many computationally expensive operations are required. The reason for this is the geometric nature of the deferral counter expiration and its dependence on the selected random backoff. The authors also extend the model to consider unsaturated conditions. However, the approach taken is based on further iterations in which the value of the transmission attempt rate is reweighted based on the offered load at every iteration until a reasonable approximation is obtained from the saturated model. As we will see in Section \ref{sec:validation}, the cost of solving this model makes it prohibitive for situations where results are required in real-time.

A similar approach to ours which also uses the renewal reward approach and that considers the backoff and deferral counters has been presented in \cite{vlachou2014performance}. However, only saturated conditions are considered. Recently, in \cite{vlachou2014new}, the same authors identify a different inaccuracy arising from the decoupling approximation under saturation conditions and a small number of contending stations, due to the high coupling between queues when the deferral counter is used. They also propose a model of the coupled system of queues based on the saturated assumption. Note that our work differs from \cite{vlachou2014new} in the following ways: \emph{i)} we evaluate the implications of the decoupling approximation on the buffers' occupancy, which can result in two different behaviours (see Section \ref{sec:demonstration}), and \emph{ii)} we propose techniques to obtain the long-term performance of the network when using a model based on the decoupling approximation under unsaturated conditions. We note that directly extending the analysis of
the coupled queues to nonsaturated conditions appears intractable.

% Other models make simplifications to the protocol

Two other analytic models of Homeplug exist in the literature (\cite{kriminger2011markov} and \cite{campista2005improving}). However, they model mechanisms that differ from the original standardised Homeplug procedure. In \cite{kriminger2011markov}, the value of the contention window is fixed for all backoff stages, while in \cite{campista2005improving} the backoff stage is incremented every time a new packet is overheard (thus having $M_i = 0,~ \forall i$). These new approaches obviously lead to a simplified analysis, as one of the dimensions of the Markov Chain is removed. However, the flexibility the standard provides by allowing to tune both the $W_i$ and $M_i$ is not captured by these frameworks.

Consequently, the analytic model in \cite{chung2006performance} has been widely used both to derive performance metrics and as a basis for extension \cite{yoon2008adaptive,koutny2011homeplug,pinero2011homeplug,koutny2013analysis,pinero2011realistic,yoon2013multichannel} as it strictly follows the procedure defined in the standard and it considers saturated as well as unsaturated conditions. This widespread use motivates our improvements in terms of \emph{i)} reducing complexity while maintaining the accuracy and retaining all features defined in the standard \cite{HomeplugStd} and \emph{ii)} understanding the transitory nature of the results predicted before saturation in \cite{chung2006performance}.

\begin{table}
\centering
\begin{tabular}{ccccc} 
Parameter & All CAs & Parameter & CA3/2 & CA1/0 \\ \hline
$M_1$ & $0$ & $W_1$ & $7$ & $7$ \\ \hline
$M_2$ & $1$ & $W_2$ & $15$ & $15$ \\ \hline
$M_3$ & $3$ & $W_3$ & $15$ & $31$ \\ \hline
$M_4$ & $15$ & $W_4$ & $31$ & $63$ \\  \hline
\end{tabular}
\caption{Parameters of the different access categories in Homeplug}\label{tbl:access_categories}
\end{table}

\section{Simplified Homeplug Analytical Model}\label{sec:model}

In this section, we present the simplified analytic model of the Homeplug MAC protocol. To this end we take a renewal reward approach \cite{kumar05,bianchi05} motivated by the fact that the attempt rate of a given node can be viewed as a regenerative process. This approach allows us to compute metrics of interest without the need to solve all state probabilities of the Markov Chain embedded in the analysis. Similar approaches have already been proposed to simplify the derivation of analytic models of the IEEE 802.11 for WLANs, see for instance \cite{medepalli2005system} and \cite{BBellalta-Eurocon2005}.  

We also apply the decoupling approximation to model the conditional (given that a packet is transmitted) collision probability independently of the backoff stage, as also done in \cite{chung2006performance} and in a large number of IEEE 802.11 analytic models (see \cite{Duffy2010} for a complete survey on this topic). Although this assumption may not be valid \cite{Huang2010}, it allows us to simplify the analysis while still accurately predicting the metrics of interest in a broad range of circumstances. The decoupling approximation is also used to model the buffer occupancy probability after a transmission as being independent of the backoff stage at which the packet was transmitted. This assumption is implicitly made in \cite{chung2006performance}. The accuracy of this assumption will be discussed in detail in Section \ref{sec:demonstration} as it is contributing factor to misprediction of results right before saturation in \cite{chung2006performance}.

The rest of assumptions and considerations taken into account are: \emph{i)} an infinite, or large enough to be considered infinite, buffer size and retry limit, \emph{ii)} exponentially distributed interarrival of packets, \emph{iii)} ideal channel conditions (i.e., packets are always received correctly in case of no collision), \emph{iv)} contention among a single access category and \emph{v)} all nodes are in mutual coverage range, that is, all nodes can overhear each other's transmissions. These are the same assumptions considered in \cite{chung2006performance} and in a number of IEEE 802.11 analytical frameworks. On one hand they allow us to compare the accuracy of our analysis and the one in \cite{chung2006performance} under the same conditions. On the other hand, they are useful to assess and understand the performance of the MAC procedure independently of other factors that may have an influence in the results in practice.

\subsection{Renewal Reward Approach}

Assuming an infinite buffer size, the mean queue occupancy ($\rho$) of a node is derived by considering the time needed to release a packet from the queue ($X$), called service time or MAC access delay, and the packet arrival rate from the network layer ($\lambda$) as $\rho=\min(\lambda X,1)$. While $\lambda$ depends on the application, the service time is computed as the sum of the following three components: \emph{i)} the total average backoff duration until the successful frame transmission, \emph{ii)} the total time on average spent in transmitting packets that result in a collision and \emph{iii)} the time spent successfully transmitting the packet:

\begin{equation}\label{eq:service_time}
X=E[w]\alpha + (n_{\rm t}-1) T_{\rm c} + T_{\rm s},
\end{equation}

where $E[w]$ denotes the average number of slots in backoff, $\alpha$ is the average slot duration and $n_{\rm t}$ is the average number of attempts to successfully transmit a packet. The duration of a successful transmission ($T_{\rm s}$) and a collision ($T_{\rm c}$) are computed as shown in Eq.~\ref{eq:Ts}. These durations account for the transmission notification intervals (PRS0 and PRS1), the time to transmit the actual frame ($T_{\rm fra}$) and the acknowledgement ($T_{\rm res}$), as well as the contention and response interframe spaces (CIFS and RIFS)\footnote{In contrast to \cite{chung2006performance}, we have considered $T_{\rm s} = T_{\rm c}$ to closely model the standard procedure. Observe that, in case of collision or frame errors, either an NACK is received or nodes wait an EIFS to provide protection from collisions for ongoing transmissions \cite{HomeplugStd}. Thus, the duration of a collision can be better approximated to that of a successful transmission.}, see \cite{HomeplugStd}.

\begin{equation}\label{eq:Ts}
T_{\rm s} = T_{\rm c} = \PRSzero + \PRSone + T_{\rm fra} + \RIFS + T_{\rm res} + \CIFS
\end{equation}

Under the decoupling assumption with an infinite number of retries, the average number of attempts to transmit a frame ($n_{\rm t}$) is computed as shown in Eq.~\ref{eq:nt}.

\begin{equation}\label{eq:nt}
 n_{\rm t}=\frac{1}{1-p},
\end{equation}

where the conditional collision probability ($p$) is obtained as the complementary of having at least one of the other $n-1$ nodes transmitting a frame in the same slot (Eq.~\ref{eq:p}), with $\tau$ denoting the attempt rate of a node. 

\begin{equation}\label{eq:p}
 p=1-(1-\tau)^{n-1}
\end{equation}

We view the attempt rate as a regenerative process, where the renewal events are when the MAC begins processing a new frame. Thus, we apply the renewal reward theorem (Eq.~\ref{eq:tau}). 

\begin{equation}\label{eq:tau}
 \tau=\frac{n_{\rm t}}{E[w]+n_{\rm t}+I}
\end{equation}

The term $I$ in Eq.~\ref{eq:tau} accounts for the number of slots in idle state (when there is no packet waiting in the queue for transmission) and is computed as the probability of having an empty queue over the probability of a packet arrival in a slot. Considering an M/M/1 queue, we then compute $I$ as in Eq.~\ref{eq:i}. %\textbf{DWM: Is this obvious? I'd have to calculate it.}

\begin{equation}\label{eq:i}
 I=\max\left(\frac{1-\rho}{1-e^{-\lambda \alpha}},0\right)
\end{equation}

The average slot duration while the node is in backoff is derived depending on the type of slot that is overheard (Eq.~\ref{eq:alpha}). A slot can be empty if no other node transmits (that occurs with $p_{\rm e}$ probability) and, in such a case, its duration is $\sigma$. Otherwise, it can be occupied due to a successful transmission (that happens with probability $p_{\rm s}$) or a collision (that occurs with $p_{\rm c}$ probability), with durations $T_{\rm s}$ and $T_{\rm c}$, respectively. 

\begin{equation}\label{eq:alpha}
 \alpha=p_{\rm s}T_{\rm s} + p_{\rm c}T_{\rm c} + p_{\rm e}\sigma
\end{equation}

Probabilities $p_{\rm s}$, $p_{\rm e}$ and $p_{\rm c}$ are obtained as follows:

\begin{align}\label{eq:p_others}
  p_{\rm s}=(n-1)\tau (1-\tau)^{n-2},\nonumber\\
  p_{\rm e}=(1-\tau)^{n-1},\nonumber\\
  p_{\rm c}=1 - p_{\rm s} - p_{\rm e}.
\end{align}

To compute $X$, it remains to obtain the value of the average number of slots spent in backoff until a packet is successfully transmitted ($E[w]$). With this aim, we follow the same approach as in \cite{chung2006performance} but we compute the average number of slots at every backoff stage. The average number of slots waiting in backoff at backoff stage $i$ (with $1 \leq i \leq m$) is shown in Eq.~\ref{eq:ewi}. This expression is obtained from \cite{chung2006performance} but it has been rearranged to account for the number of backoff slots instead of the total delay to transmit a packet. The reader is referred to \cite{chung2006performance} for details.

\begin{figure*}[!t]
\begin{equation}\label{eq:ewi}
\begin{split}
 E[w_i] = \frac{1}{W_i+1} \Bigg[ \frac{M_i^2+M_i}{2} &+ \sum_{k=1}^{W_i-M_i} \Bigg\{ \sum_{l=0}^{k-1} \binom{M_i+l}{l} (1-p_{\rm b})^l p_{\rm b}^{M_i} p_{\rm b} (l+M_i+1) \\
	  & + \sum_{l=0}^{M_i} \binom{M_i+k}{k+l} (1-p_{\rm b})^{k+l} p_{\rm b}^{M_i-l} (k+M_i) \Bigg\} \Bigg]
\end{split}
\end{equation}
\end{figure*}

In Eq. \ref{eq:ewi}, $p_{\rm b}$ refers to the probability of overhearing another transmission while the node is in backoff:

\begin{equation}\label{eq:pb}
 p_{\rm b}=1-(1-\tau)^{n-1}.
\end{equation}

A node moves to the next backoff stage (if it has not reached its maximum value) whenever it fails to transmit a packet. We define a failure here as the case in which either: \emph{i)} the backoff expires leading to the transmission of the packet and there is a collision or \emph{ii)} the deferral counter expires so that the node defers its transmission. The probabilities of each of these situations depend on the backoff stage as they are a function of the value of $W_i$ and $M_i$. Thus, we compute the probability of failure at backoff stage $i$ as shown in Eq.~\ref{eq:pf}. 

\begin{equation}\label{eq:pf}
p_{\rm f}^{(i)} =p \cdot p_{\rm bo}^{(i)} + p_{\rm defer}^{(i)},
\end{equation}

where $p_{\rm bo}^{(i)}$ is the probability that the backoff expires in backoff stage $i$. On the other side, $p_{\rm defer}^{(i)}$ denotes the probability that the deferral counter reaches zero and a new packet is overheard in the given backoff stage. This probability is also obtained from \cite{chung2006performance} and it is calculated as:

\begin{equation}\label{eq:pdefer}
 p_{\rm defer}^{(i)} = \frac{1}{W_i+1} \sum_{k=1}^{W_i-M_i} \sum_{l=0}^{k-1} \binom{M_i+l}{l} (1-p_{\rm b})^l p_{\rm b}^{M_i} p_{\rm b}.
\end{equation}

The probability that the backoff expires is just the complementary of the probability to defer (Eq.~\ref{eq:pbo}).

\begin{equation}\label{eq:pbo}
p_{\rm bo}^{(i)} = 1 - p_{\rm defer}^{(i)}
\end{equation}

Thus, the total average number of slots in backoff is obtained by summing the number of slots waiting at each stage weighted by the probability of moving to that backoff stage (see Eq.~\ref{eq:ew}). Note that the probability to move to a given backoff stage is the probability to face a failure in the previous ones. An extra term is considered for the last backoff stage as the node returns to it after every failure. Assuming a packet is retransmitted until it is successfully transmitted the last term follows.

\begin{equation}\label{eq:ew}
\begin{split}
 E[w] & = E[w_1] + \sum_{i=2}^{m-1} E[w_i] \prod_{j=1}^{i-1} p_{\rm f}^{(j)} \\
      & + E[w_m] \prod_{j=1}^{m-1} p_{\rm f}^{(j)} \frac{1}{1-p_{\rm f}^{(m)}}
\end{split}
\end{equation}

Finally, having the payload length ($L$), we obtain the throughput as:

\begin{equation}\label{eq:S}
S =  \rho \frac{L}{X}
\end{equation}

The previous expressions can be solved by using an iterative numerical method. If compared to the previous analytic model in \cite{chung2006performance}, the renewal reward approach allow us to reduce the complexity by removing two iteration loops: the one involving the calculation of the state probabilities of the Markov Chain as well as the one regarding the total average MAC access delay computation.

\subsection{Precomputation of $p_{\rm defer}^{(i)}$ and $E[w_i]$}

% just a function of pb, W and M, linear extrapolation

The most computationally expensive operations of our simplified analytic model are the computation of $p_{\rm defer}^{(i)}$ and $E[w_i]$ (Eq.~\ref{eq:ewi} and \ref{eq:pdefer}, respectively). However, both are a function of just three variables: $p_{\rm defer}^{(i)}=f(W_i, M_i,p_{\rm b})$ and $E[w_i]=g(W_i, M_i,p_{\rm b})$. Since $W_i$ and $M_i$ are fixed parameters and $p_{\rm b} \in [0,1]$, we can easily precompute the values of these two metrics for each relevant value of $W_i$, $M_i$ and $p_{\rm b}$.

The error of the approximated results can be reduced by applying a linear interpolation between the two closest precomputed values of $p_{\rm b}$. Results presented in this work have been obtained using the precomputed values as they have been found indistinguishable (using a step size equal to $10^{-4}$ for $p_{\rm b}$) from the ones obtained using Eq.~\ref{eq:ewi} and \ref{eq:pdefer}. %\textbf{DWM: Should we say the step size used?}

\subsection{Exponential Approximation to $p_{\rm defer}^{(i)}$}

We consider a further optional step in reducing the complexity of the analytic model. We may calculate the probability to defer a packet in a given backoff stage by approximating both the backoff countdown and the deferral counter with exponential distributions with means chosen to match the actual distributions. We approximate the backoff counter and the deferral counter exponentially distributed with rate $\beta_i=2/W_i$ and $\gamma_i=p_{\rm b}/(M_i+1)$, respectively. Then, we can easily derive the probability to defer as the probability that the deferral counter expires first, that is:

\begin{equation}\label{eq:pdefer_exp_approx}
p_{\rm defer}^{(i)}=\frac{\gamma_i}{\beta_i+\gamma_i}.
\end{equation}

\begin{figure}[t!!!!!]
\centering
\subfigure[CA3/2]{\includegraphics[width=2.6in]{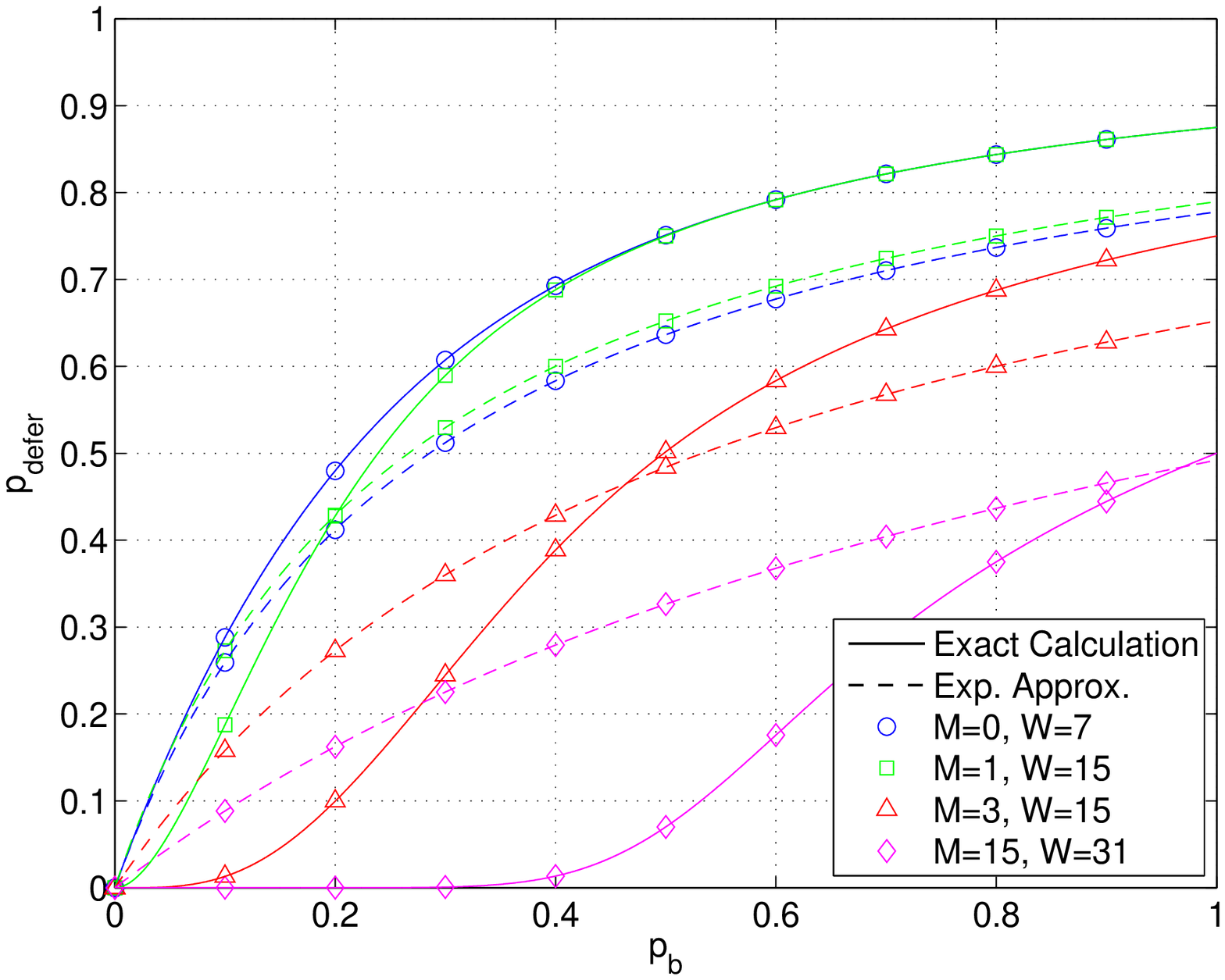}\label{fig:exp_approx_ca3ca2}}
\subfigure[CA1/0]{\includegraphics[width=2.6in]{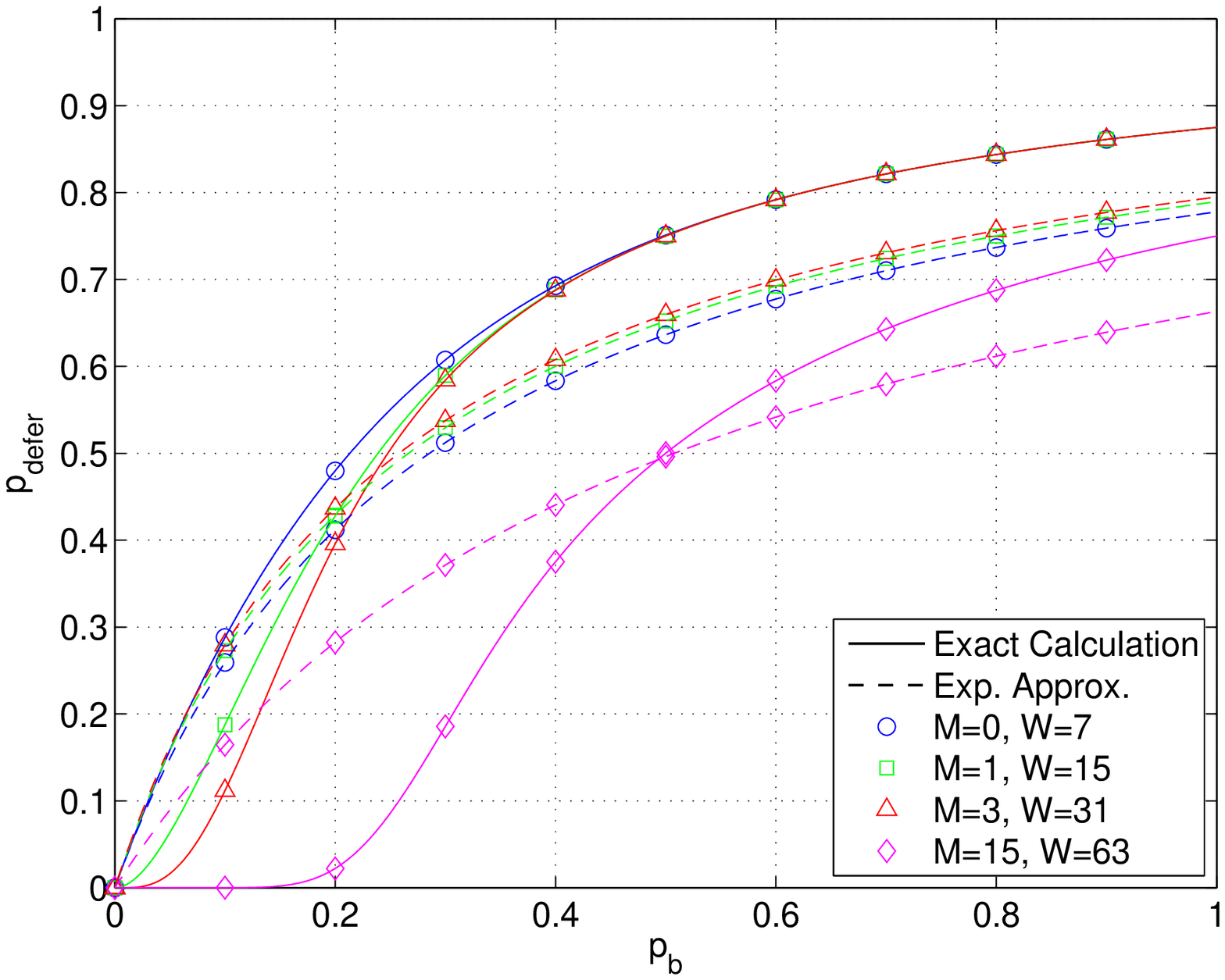}\label{fig:exp_approx_ca1ca0}}
\caption{Comparison of $p_{\rm defer}^{(i)}$ computed following the exact calculation and the exponential approximation.}
\label{fig:exp_approx_comp}
\end{figure}

Although the otional approximation considerably reduces the complexity of the calculation of $p_{\rm defer}^{(i)}$ in Eq.~\ref{eq:pdefer}, it only provides a rough estimate as shown in Fig.~\ref{fig:exp_approx_ca3ca2} (CA3/2) and \ref{fig:exp_approx_ca1ca0} (CA1/0) for an arbitrary value of $p_{\rm b}$. The disagreement is more notable for high values of $M_i$. The reason for this result is the inability of the approximation to model the cases in which the random backoff selected at stage $i$ is smaller than $M_i$. In such a case, regardless of the value of $p_{\rm b}$, the backoff counter will always expire first, yet the exponential approximation always gives some probability that either expire first. However, note that the estimation is accurate for certain values of $p_{\rm b}$, $M_i$ and $W_i$. As will be shown in Section \ref{sec:validation}, the approximation does provide a good estimate of throughput and delay for certain ranges of $p$ and thus, could be safely used in the cases identified.

\section{On the Performance Results Right Before Saturation}\label{sec:demonstration}

Before the performance evaluation, in this section, we demonstrate that the results presented in \cite{chung2006performance} just prior to predicted saturation correspond to a transitory phase of the network. To do so we refer to previous literature regarding the decoupling approximation and to show that the coupled system of queues under infinite buffer size is unstable in this regime. Thus, with an infinite queue size, higher throughput than the one found in saturation, as predicted in \cite{chung2006performance}, cannot be maintained in the long term. Furthermore, we support this demonstration by experimental evaluation in which we show, under certain circumstances, the extremely long time during which the network remains in the transitory period before moving to the long-term behaviour.

\subsection{The Decoupling Approximation}

The decoupling approximation allows to make the analysis tractable by assuming that each queue can be modelled independently of the dynamics of the rest of queues in the network. However, one has to be careful when interpreting the results obtained. As already observed in \cite{Duffy2010}, unsaturated analytic models that decouple the queue dynamics can provide two different solutions in certain regimes. In particular, when the packet arrival rate is slightly higher than the maximum load that the system could serve in saturation. These analytic models do not consider the number of instantaneous contending stations, i.e., the number of queues that have at least a packet buffered at the same time. Neglecting this fact makes it impossible to identify the actual regime at which the queues operate in this region. This is caused by the possibility of facing two extreme cases: the queues being mostly empty or saturated conditions. On one hand, if a reduced number of nodes are contending for the channel, the channel access delay is small, thus, making unlikely that a high number of packets accumulate in the buffer for transmission. This implies higher throughput than found in saturation, as the conditional collision probability is reduced. On the other hand, if a higher number of nodes are contending, the channel access delay increases and so does the probability of having more packets accumulating for transmission. The effect of the latter case, under infinite buffer size, is the saturation of the buffer. Analytical models that do not consider the number of instantaneous contending stations are not always able to differentiate among the two different solutions and thus, iterative algorithms may converge to either, depending on the starting parameters.

\subsubsection{Which solution corresponds to the performance of the network in the long term?}

As already conjectured in \cite{Duffy2010}, when considering infinite buffer size, the lowest throughput solution is the one that corresponds to the long-term behaviour of the network, while the other solution is just transitory. This has been demonstrated in \cite{borst2008stability}, where a coupled system of parallel queues with infinite buffer size and state-dependent service rates is analysed. The system of queues is found to be unstable (the stability limit is surpassed) when the packet arrival rate ($\lambda$) is higher than the service rate that can be achieved when all nodes are simultaneously contending for the channel. This service rate corresponds to the one found in saturation. Therefore, when the packet arrival rate is above the service rate in saturation, all queues will eventually have an increasing number of packets. Thus, the long-term performance of the network corresponds to that in saturation (all nodes will have at least a packet buffered for transmission). The results presented in \cite{chung2006performance} show higher throughput than the one found in saturation when the stability limit is slightly surpassed. However, given that the queues are unstable in that regime, the long-term throughput should correspond to the one in saturation instead.   

%Observe that, the instability of the queues translates to the case in which all queues have at least a packet to transmit. Thus, from a networking point of view, the solution that gives the saturation throughput is the one that corresponds to the long-term performance of the network. Therefore, the region just before the saturation point identified in \cite{chung2006performance}, showing higher throughput than the one found in saturation, is actually just \emph{after} the stability limit of the queues and. Hence, the long-term throughput should correspond to the one in saturation instead.

\begin{figure}[!hhhhtb]
\centering
\includegraphics[width=2.6in]{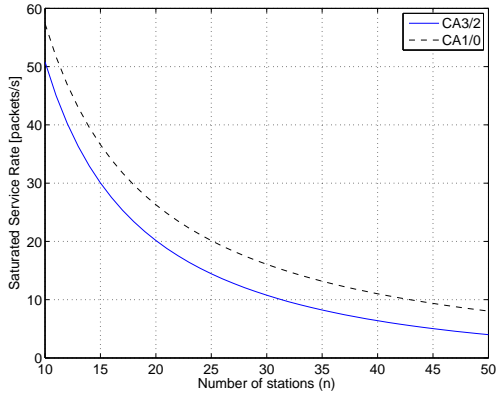}
\caption{Service rate in saturated conditions.}
\label{fig:mu_sat}
\end{figure}

In order to identify the stability limit regime, Fig.~\ref{fig:mu_sat} shows, for different number of nodes contending for the channel and access categories, the service rate obtained in saturation conditions. The parameters used are shown in Table \ref{tbl:parameters} and correspond to those in \cite{chung2006performance}, which are defined in Homeplug 1.0 \cite{HomeplugStd}. Note that the service rate in saturated conditions is just $\mu_{\rm sat} = 1/X_{\rm sat}$, where $X_{\rm sat}$ denotes here the service time considering that all nodes always have a packet ready for transmission. From \cite{borst2008stability} (for a homogeneous system), we identify the stability condition of the system of coupled queues as $\lambda < \mu_{\rm sat}$. As will be shown in Section \ref{sec:validation}, when the stability condition of the network is not satisfied, there is a potential for obtaining two solutions from the analytic model. The stability limit is surpassed if $\lambda$ is higher than the service rate depicted. It is important to emphasise that the potential for obtaining two solutions decreases when $\lambda >> \mu_{\rm sat}$ as the probability of having a large number of the nodes simultaneously contending for the channel increases.

When using iterative numerical solvers, setting certain starting parameters can provide a particular solution among the two possible ones. Setting the metrics assuming a highly-loaded network starts the numerics closer to the saturated solution. On the contrary, assuming lightly-loaded conditions, starts the iterative loop closer to the transitory phase. When there is only one solution, the solver converges to it independently of starting under any of these initial conditions. We will use this technique in Section \ref{sec:validation}, to obtain both solutions from the analysis. This technique, although useful to distinguish the two solutions, is not able to differentiate them when both are very close. However, we know that after surpassing the stability limit (as described above) the throughput that should be obtained is the one corresponding to saturation. 

\subsection{The Long Transitory Phase in Experimental Evaluation}

When $\lambda >> \mu_{\rm sat}$, the system rapidly moves to the long-term behaviour. However, performing an experimental evaluation right after the stability limit can provide wrong results since, as we have found in this work, the length of the transitory phase can be extremely long. We have found that if the experiments are started with the queues empty, it can take a long time to reach the long-term performance since the system has to reach a point at which a large number of nodes are simultaneously contending for the channel and start to have their queues filled with an increasing number of packets. The possibility of a long transitory phase for random MAC protocols in this regime was postulated in \cite{suleiman2008impact}, however no experimental findings or proof was provided. Here we take the step of showing experimentally that for Homeplug MAC this is exactly the case. 

For that purpose we have run long simulations with load just above the stability limit with queues long enough to be considered infinite ($1000$ packets). From the simulations we have measured the throughput and the queue size (maximum, average and minimum) of all the nodes in the network in $1$s-long time intervals. Simulation results are obtained using a custom simulator based on the SENSE framework \cite{chen2004sense}. We have considered the case in which $50$ nodes with packets belonging to CA3/2 access categories are accessing the channel, as larger numbers of nodes help us highlight the discrepancy between the previously presented results (see \cite{chung2006performance} Fig.~6 and 7) and the saturated solution. We begin the simulations with the queues empty.

\begin{figure*}[th!!!!!]
\centering
%\subfigure[$\lambda=8.5$ packets/s]{\includegraphics[width=2.5in]{./Figures/temporal_tpinst_n_50_lambda_8dot5}}
%\subfigure[$\lambda=8.5$ packets/s]{\includegraphics[width=2.6in]{./Figures/temporal_q_zoom_n_50_lambda_8dot5}}\\

\subfigure[$\lambda=8$ packets/s]{\includegraphics[width=2.5in]{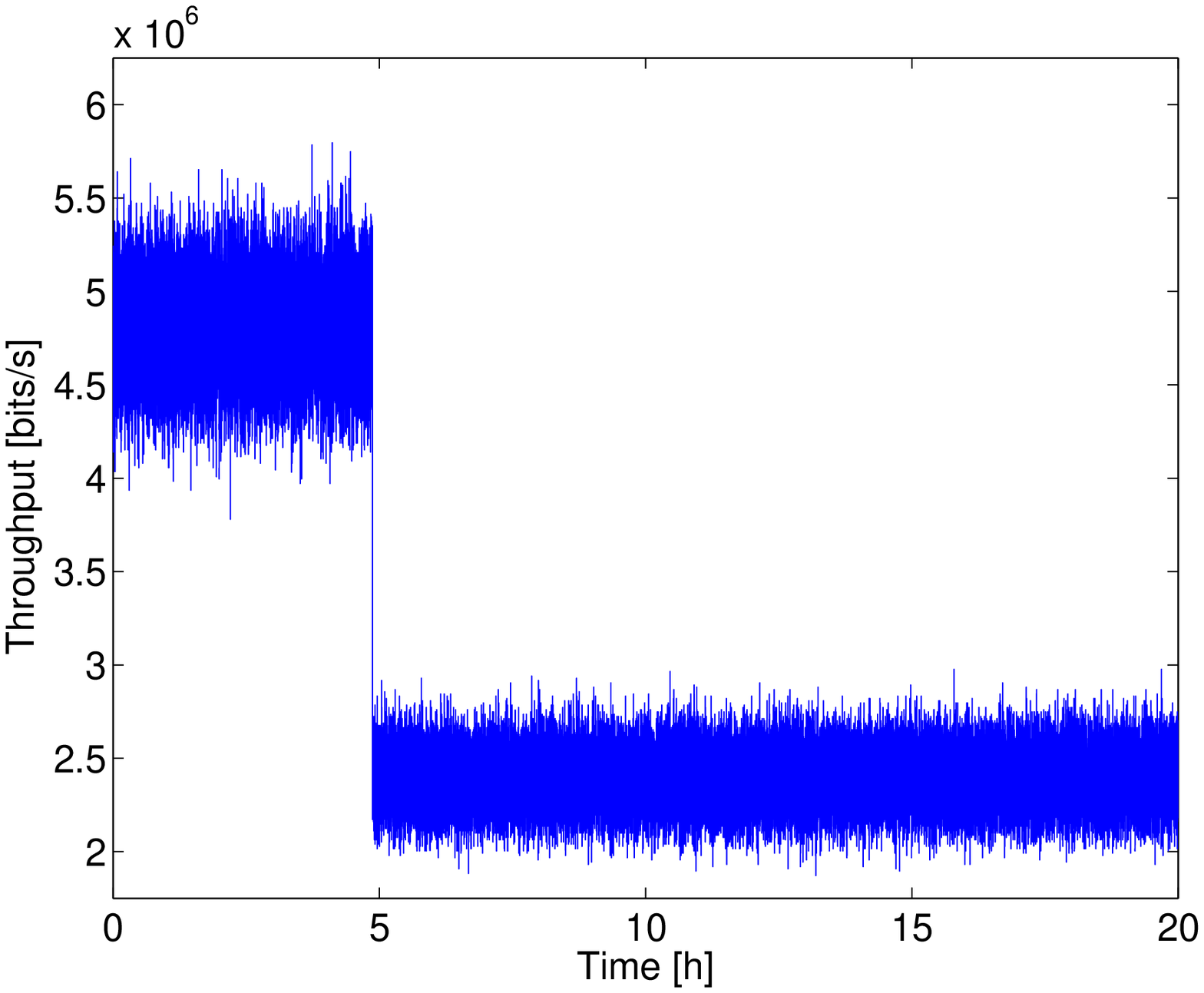}}
\subfigure[$\lambda=8$ packets/s]{\includegraphics[width=2.55in]{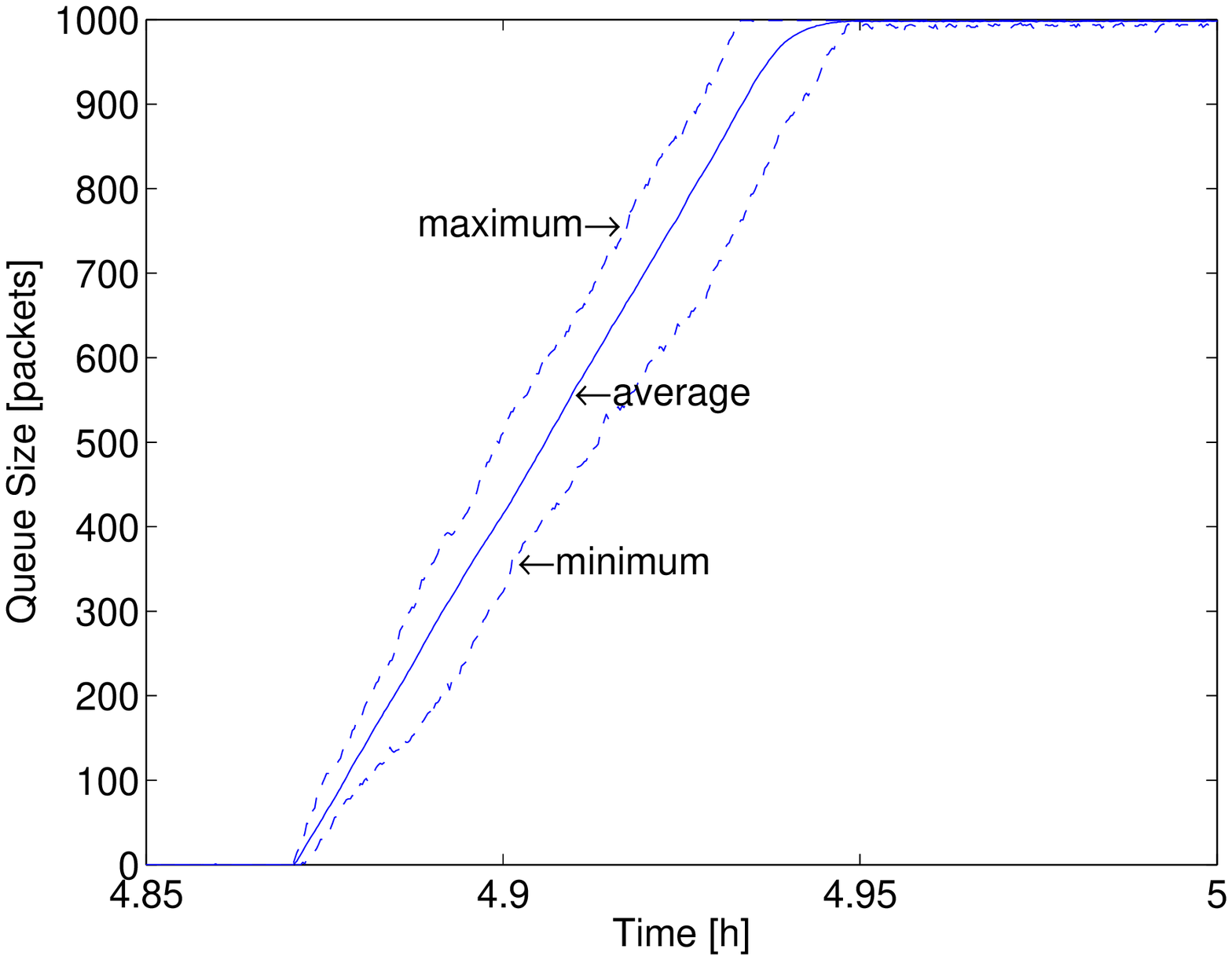}}\\

\subfigure[$\lambda=7.5$ packets/s]{\includegraphics[width=2.5in]{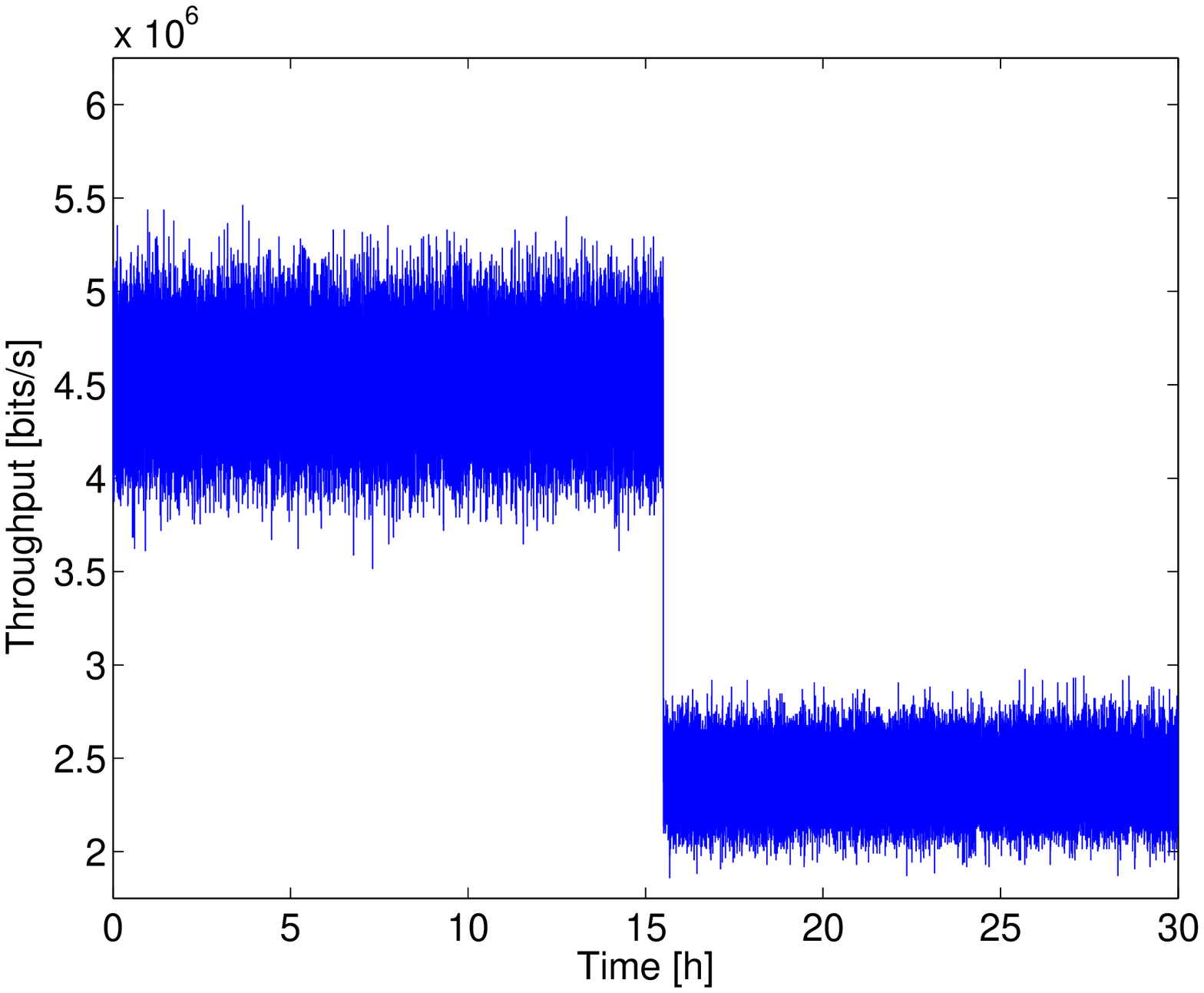}}
\subfigure[$\lambda=7.5$ packets/s]{\includegraphics[width=2.6in]{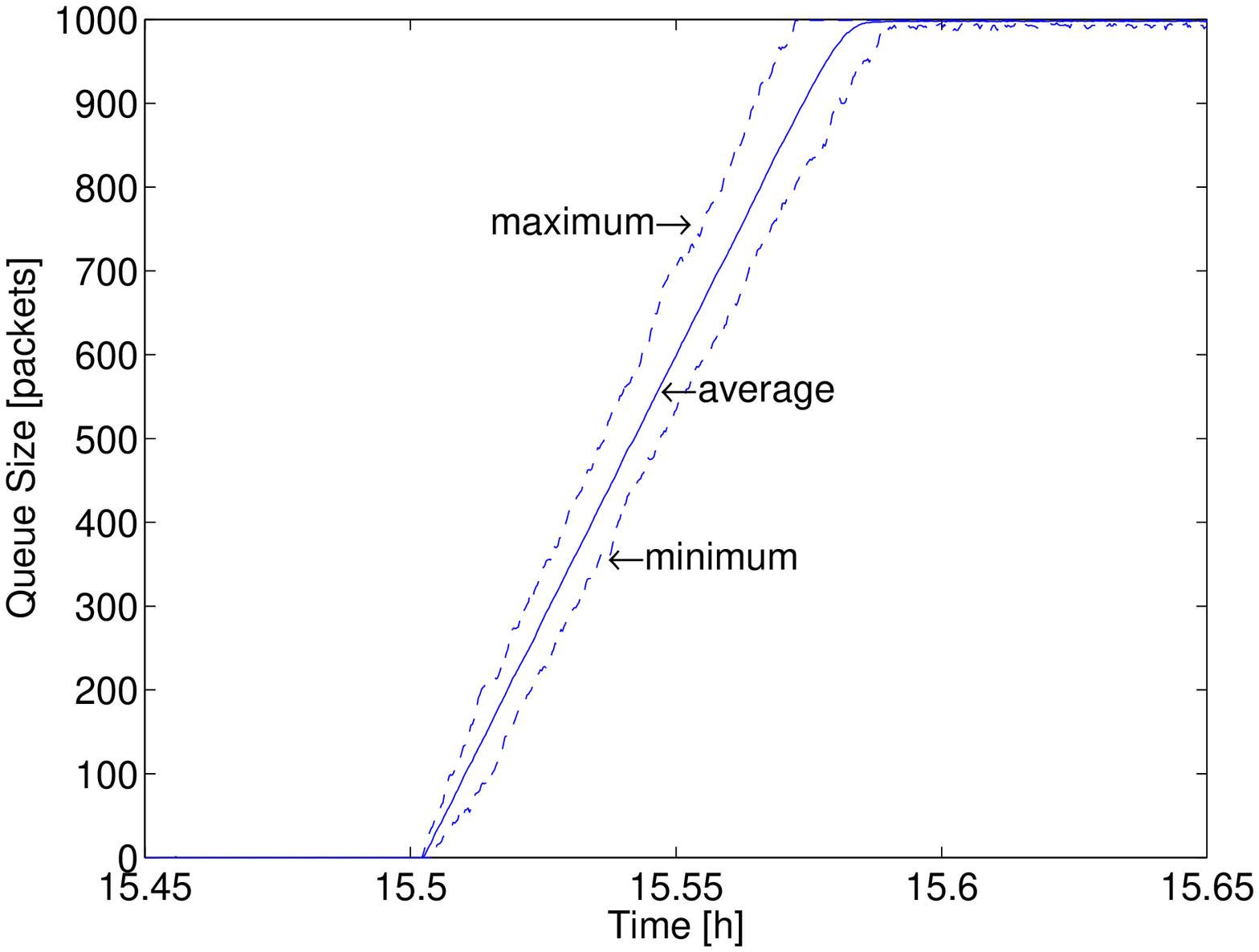}}
\caption{Evolution of throughput and queue size (CA3/2, $n=50$ nodes).}
\label{fig:temporal}
\end{figure*}

Observe in Fig.~\ref{fig:temporal}(a,c) that, for different values of the packet arrival rate, there is a sharp change from the transitory behaviour to the long-term solution. In Fig.~\ref{fig:temporal}(b,d), we zoom in on the queue behaviour around this sharp change. We see that the queues start to fill with packets until the maximum size is reached. The conjecture that the transitory phase is of long duration is also supported, as we observe a duration of the order of hours in the runs shown.

It is important to emphasise that the results presented in Fig.~\ref{fig:temporal} are each from a single simulation run. The time at which the change in behaviour occurs varies in different runs due to its stochastic nature. To characterise the length of the transitory period one must perform analysis considering the coupled dynamics of the system of queues. Due to its difficulty, especially considering the complexity of the Homeplug MAC access procedure, we consider this analysis out of the scope of this work.

Because of this long transitory period, care is required in designing the simulations. One way to obtain the long-term performance is to start with the queues empty, run the simulation for a long time until the queues are saturated and then start taking the statistics of the performance metrics of interest. However, we suggest starting the simulations with a number of packets preloaded in the queues as a more practical way to force the system to enter into the long-term operating state. If the queues are stable, nodes will be able to release these preloaded packets in a reasonable amount of time. If the queues are unstable, we have started the simulation closer to their stationary regime, and will see the long-term throughput more quickly. We use this technique to avoid the transitory period in the next section in order to suppress the effect of the transitory on performance results.

\section{Performance Evaluation}\label{sec:validation}

In this section, we present the validation of our analytic model. The validation is divided according to saturated and unsaturated conditions. The later also show the two different solutions the analytic model may provide as well as the long-term performance to which the system converges to, obtained from experimental evaluation. We then evaluate the effect of the deferral counter on misprediction errors and, in the last subsection, we show the complexity reduction of our simplified analysis.

Simulations are performed, as in the last section, using the SENSE framework and parameters shown in Table \ref{tbl:parameters}. Results show averages from simulation runs of $10000$~s.

\begin{figure*}[ht!!!!!!!!!!]
\centering
\subfigure[Throughput]{\includegraphics[width=2.5in]{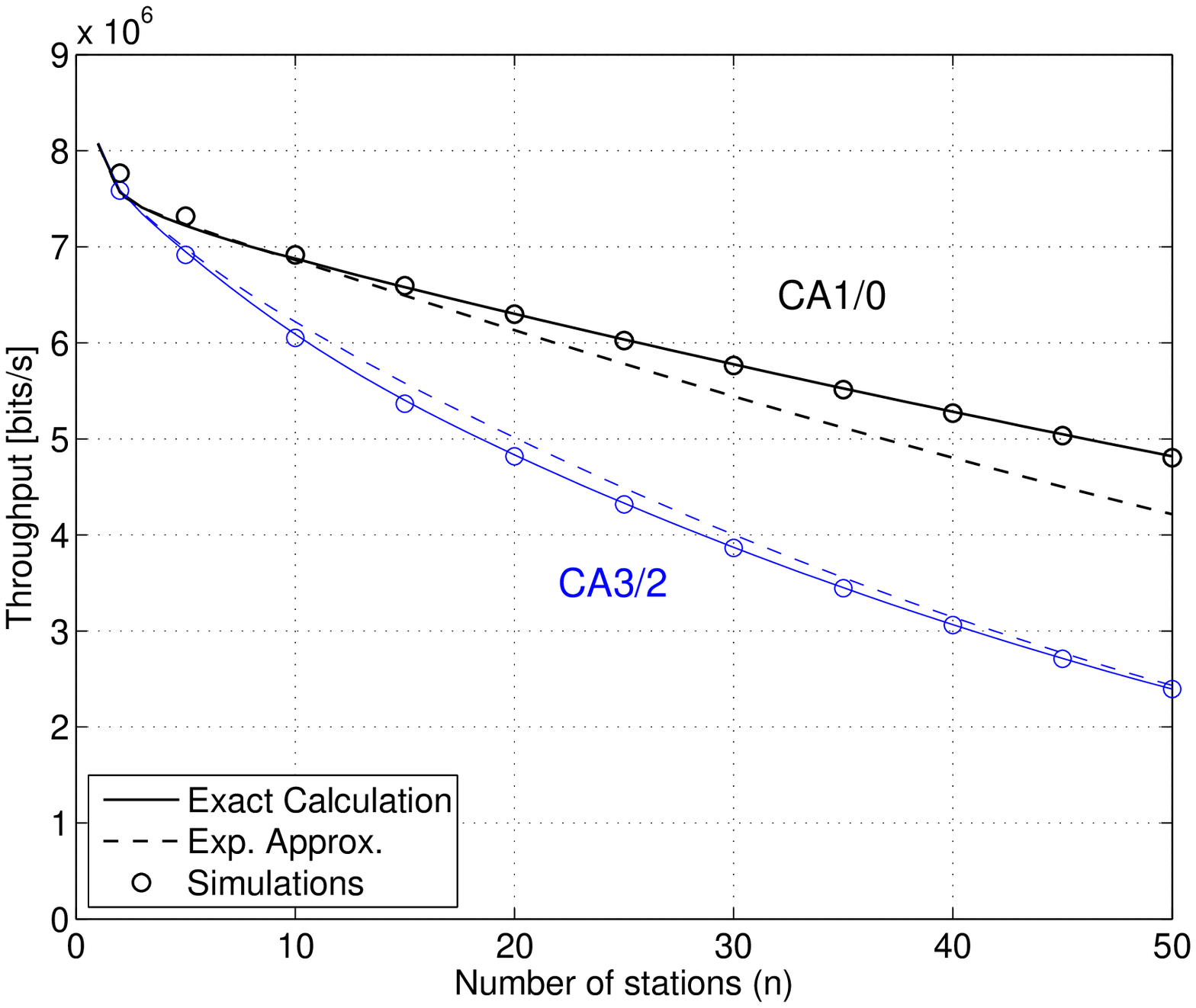}\label{fig:sat_s}}
\subfigure[MAC access delay]{\includegraphics[width=2.6in]{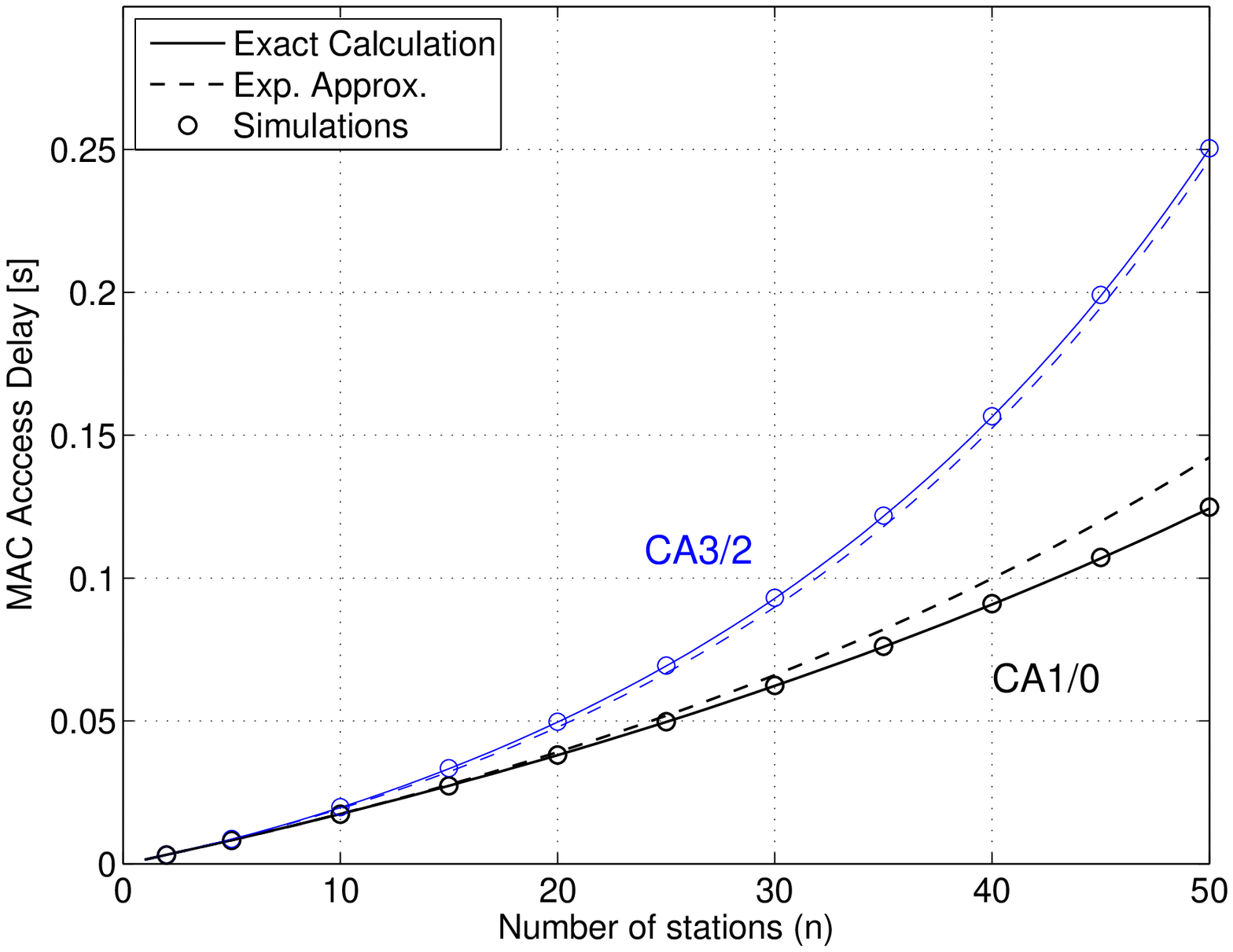}\label{fig:sat_d}}
\caption{Performance results in saturation conditions. Comparison among the exact calculation, the exponential approximation of $p_{\rm defer}^{(i)}$ and simulations.}
\label{fig:sat}
\end{figure*}

\begin{table}
\centering
\begin{tabular}{cc} 
Parameter & Value in Homeplug 1.0\\ \hline
Data rate ($R$) & $14$~Mbps \\ \hline
Frame transmission time ($T_{\rm fra}$)/$L$ & $1153.5$~$\mu$s/$1500~$bytes \\\hline
ACK transmission time ($T_{\rm res}$) & $72$~$\mu$s \\ \hline
Slot time ($\sigma$) & $35.84$~$\mu$s \\ \hline
Data-ACK interframe space ($\rm{RIFS}$) & $26$~$\mu$s \\\hline
Contention interframe space ($\rm{CIFS}$) & $35.84$~$\mu$s \\ \hline
Tx. indication slots ($\PRSzero = \PRSone$) & $35.84$~$\mu$s \\ \hline
\end{tabular}
\caption{Parameters Homeplug 1.0}\label{tbl:parameters}
\end{table}

\subsection{Saturated Conditions}

Throughput and channel access delay metrics in saturation conditions for different number of nodes are shown in Fig.~\ref{fig:sat_s} and \ref{fig:sat_d}, respectively. Results from \emph{i)} the exact computation (obtained from precomputing the values of $p_{\rm defer}^{(i)}$ and $E[w_i]$), \emph{ii)} the exponential approximation to $p_{\rm defer}^{(i)}$ and \emph{iii)} simulations are depicted. Observe that the results obtained from the exact calculation are in agreement with the performance predictions and simulations presented in \cite{chung2006performance} in Figs. 4 and 5 (slight differences appear due to our $T_{\rm s} = T_{\rm c}$ consideration). Moreover, we also see how the optional approximation is able to accurately predict both throughput and channel access delay in all the cases considered for access categories CA3/2. For access categories CA1/0, the exponential approximation is accurate for a small number of contending nodes (less than $15$). For a large number of nodes, the approximation 
provides a rough estimate. 

\begin{figure}[!tb]
\centering
\includegraphics[width=2.6in]{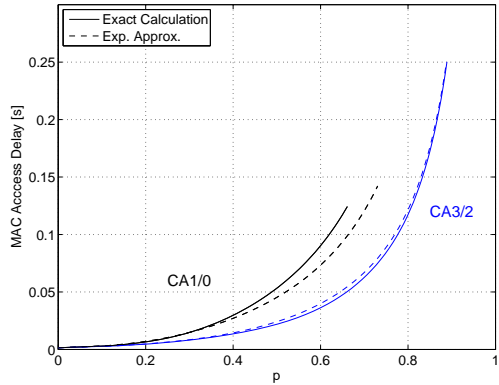}
\caption{Channel access delay vs. $p$.}
\label{fig:sat_d_vs_p}
\end{figure}

To study when the exponential approximation is useful, Fig.~\ref{fig:sat_d_vs_p} shows the channel access delay vs. the conditional collision probability, obtained using the exact calculation and the exponential approximation. This plot allows us to identify when the exponential approximation gives a good estimate, independent of the exact configuration, since the channel access delay as a function of the conditional collision probability is independent of the number of nodes contending \cite{malone2007modeling}. Thus, we observe that the exponential approximation is accurate for access categories CA3/2 for all configurations, while for access categories CA1/0 it provides a good estimate when $p<0.4$.

\subsection{Unsaturated Conditions}

\begin{figure*}[!tb]
\centering
\subfigure[Throughput CA3/2]{\includegraphics[width=2.5in]{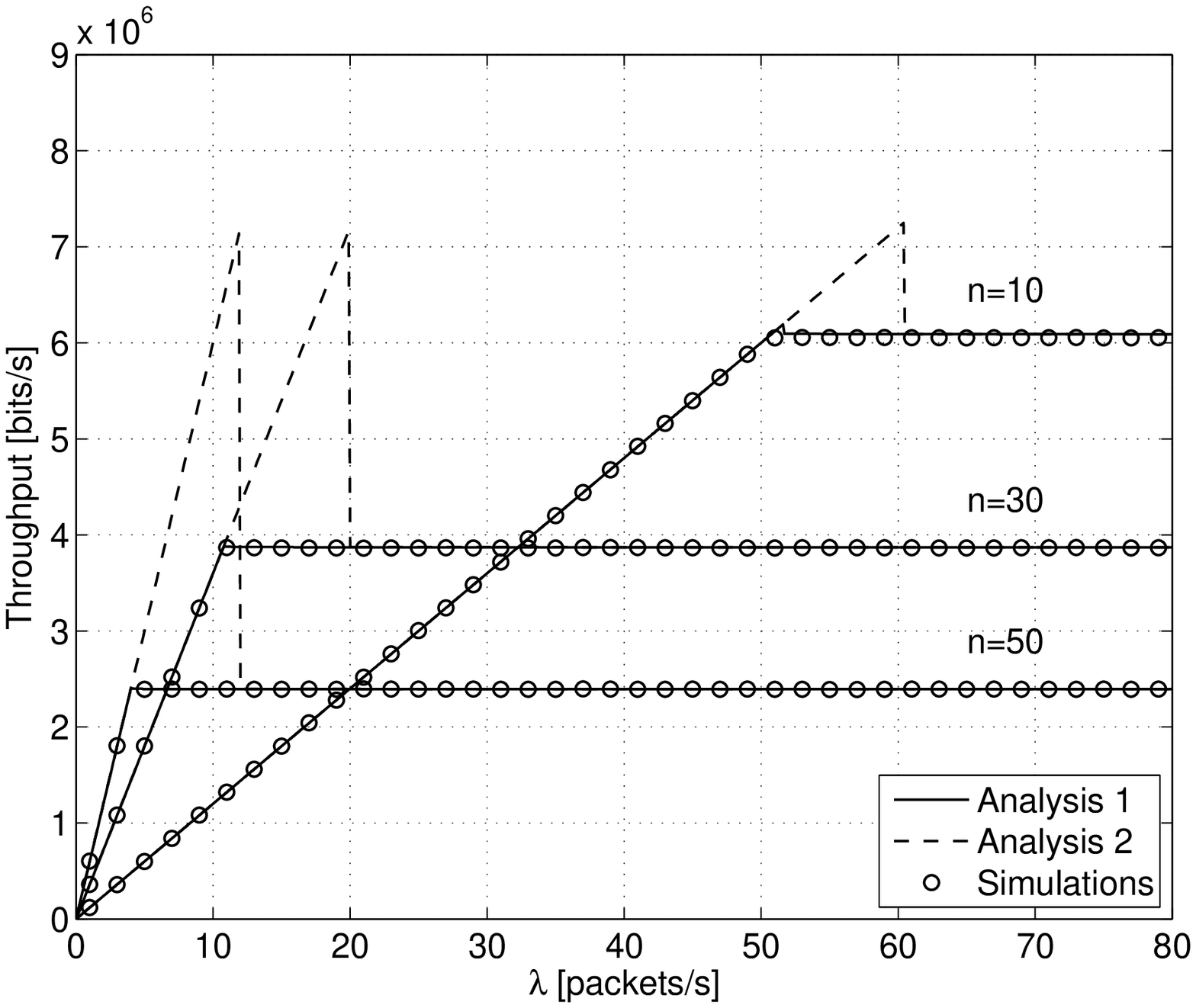}\label{fig:unsat_s_ca3ca2}}
\subfigure[MAC access delay CA3/2]{\includegraphics[width=2.6in]{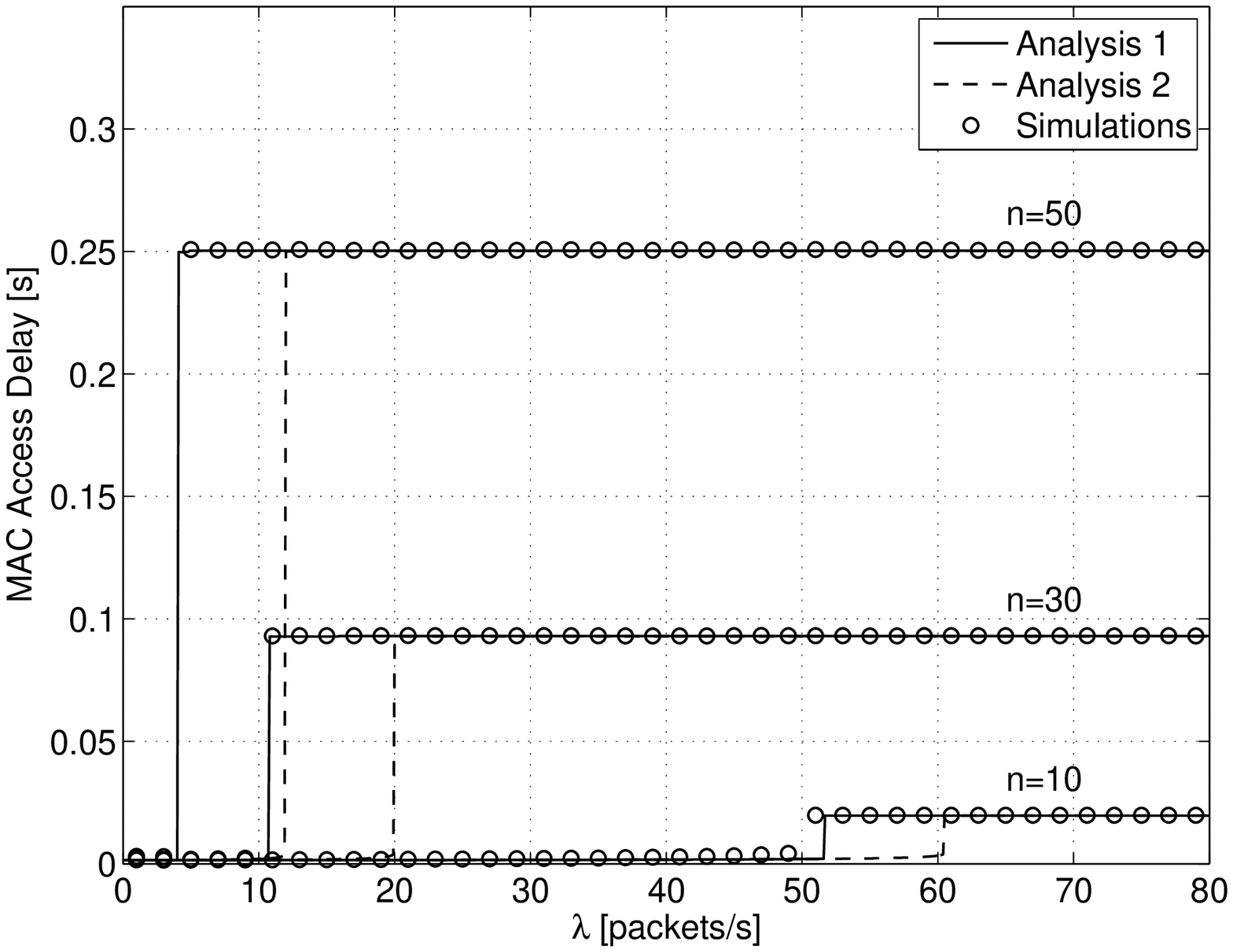}\label{fig:unsat_d_ca3ca2}}\\
\subfigure[Throughput CA1/0]{\includegraphics[width=2.5in]{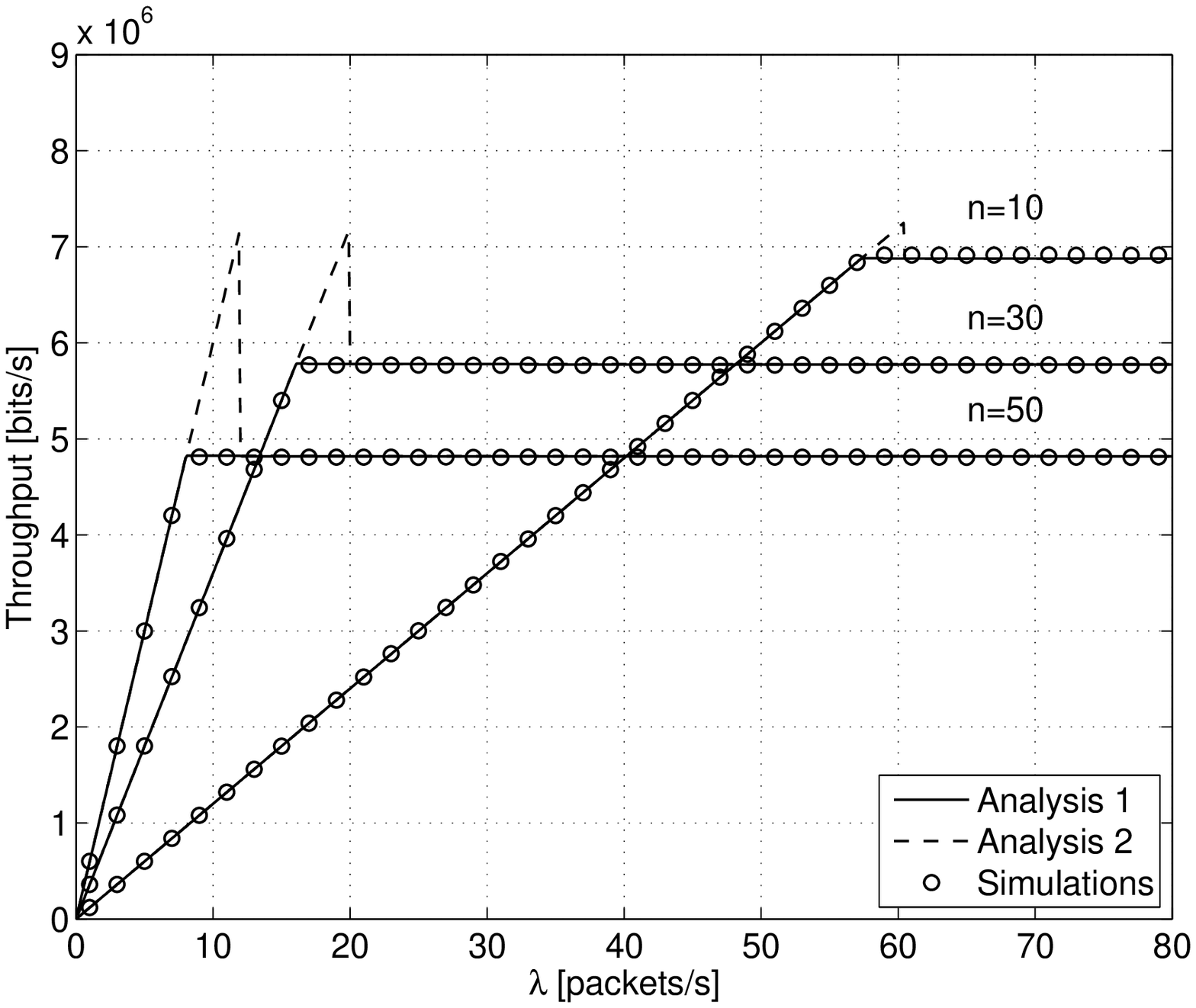}\label{fig:unsat_s_ca1ca0}}
\subfigure[MAC access delay CA1/0]{\includegraphics[width=2.6in]{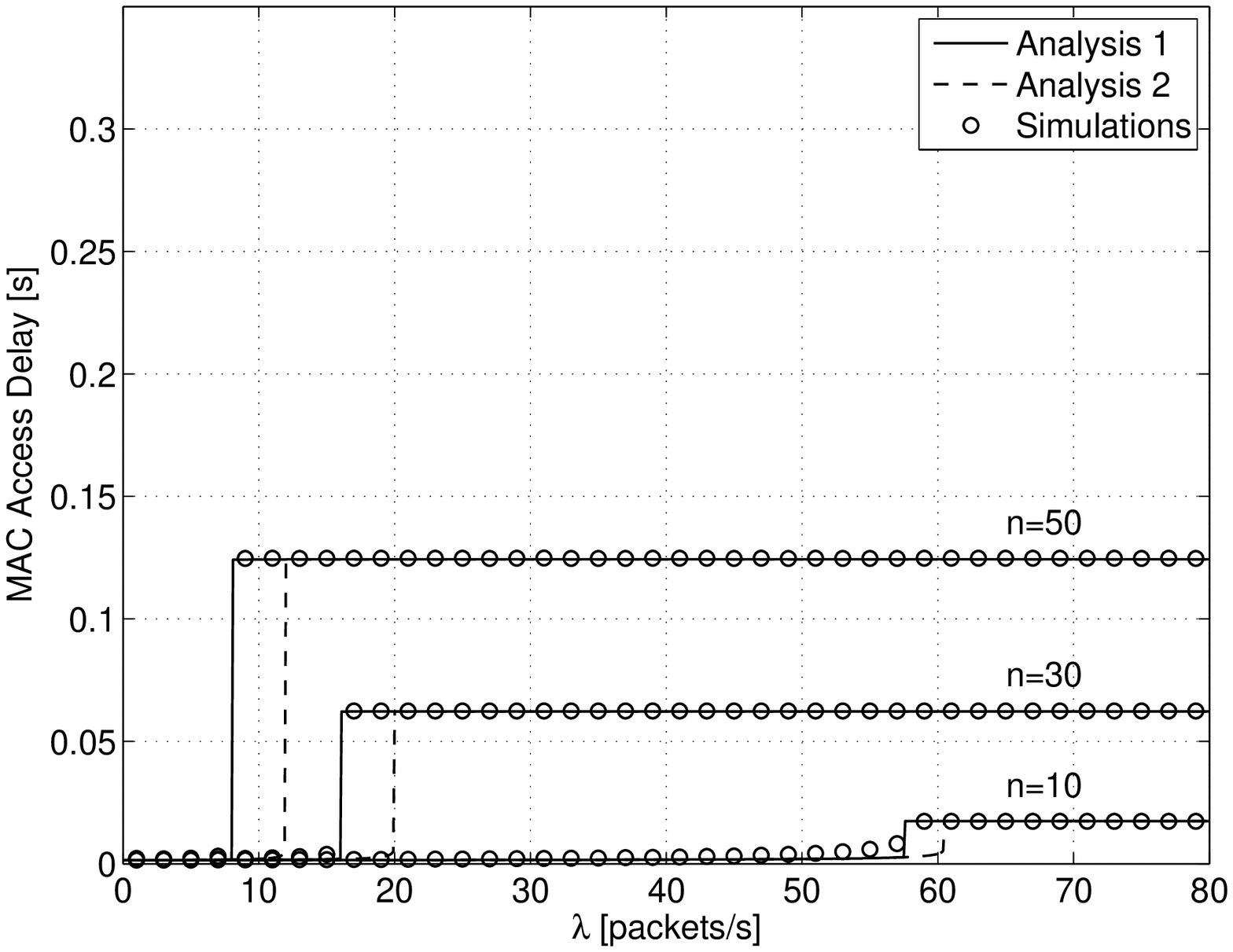}\label{fig:unsat_d_ca1ca0}}
\caption{Performance results in unsaturated conditions. Comparison among the two solutions derived from the exact analysis and simulations.}
\label{fig:unsat}
\end{figure*}

Results in unsaturated conditions for $n=10,30$ and $50$ and varying the packet arrival rate $\lambda$ are shown in Fig.~\ref{fig:unsat}. Values of throughput and channel access delay for the different access categories obtained from the exact computation and simulations are depicted. 

For the analytic results, we show two solutions, corresponding to beginning the numerics for solving the model with $I = 0$ slots (no idle periods between a departure and the next packet arrival, labelled \emph{Analysis 1}) or $I=1000$ slots (long idle periods between departures and next packet arrivals, labelled \emph{Analysis 2}). Observe in Fig.~\ref{fig:unsat} that the model may converge to different solutions. As expected, just after the stability limit, the analytic model provides two different solutions depending on the starting values we provide to the iterative loop. We see that \emph{Analysis 1} provides the lowest throughput, while \emph{Analysis 2} shows higher throughput.

We use here the technique of preloading the queues as already discussed in the last section. Thus, we have averaged the values of $10000$~s-long simulations with the queues preloaded with $50$ packets. We have compared results from \emph{Analysis 1} and find agreement with the long-term performance obtained after the transitory phase. 

Note how, excluding the results right after saturation (as we already discussed), the results obtained are also in agreement with the outcomes obtained in \cite{chung2006performance} in Figs. 6-9, validating, as in the saturated case, the accuracy of our proposed analysis.

\begin{figure*}[!tb]
\centering
\subfigure[Throughput CA3/2]{\includegraphics[width=2.5in]{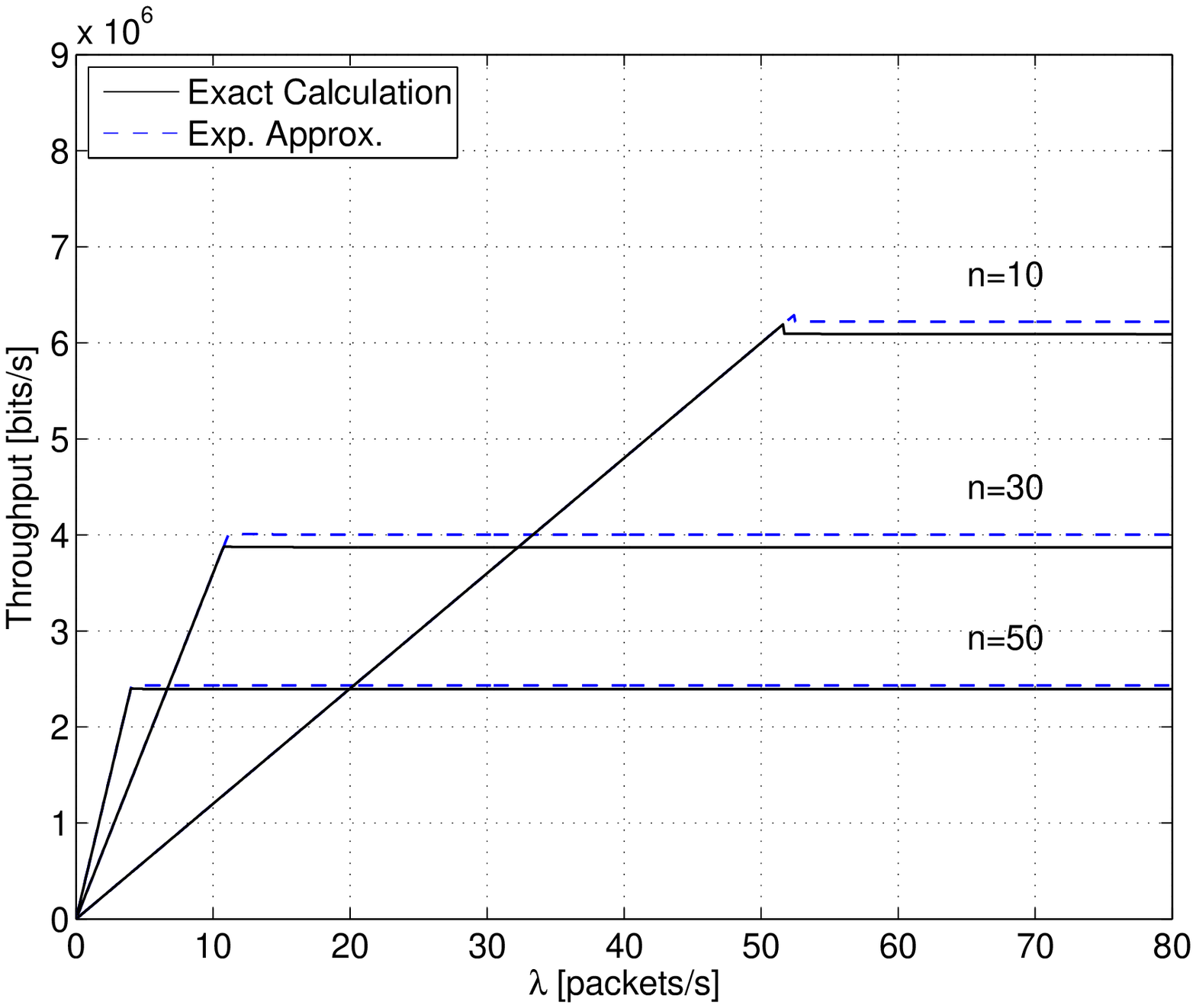}\label{fig:unsat_s_ca3ca2_exp}}
\subfigure[MAC access delay CA3/2]{\includegraphics[width=2.6in]{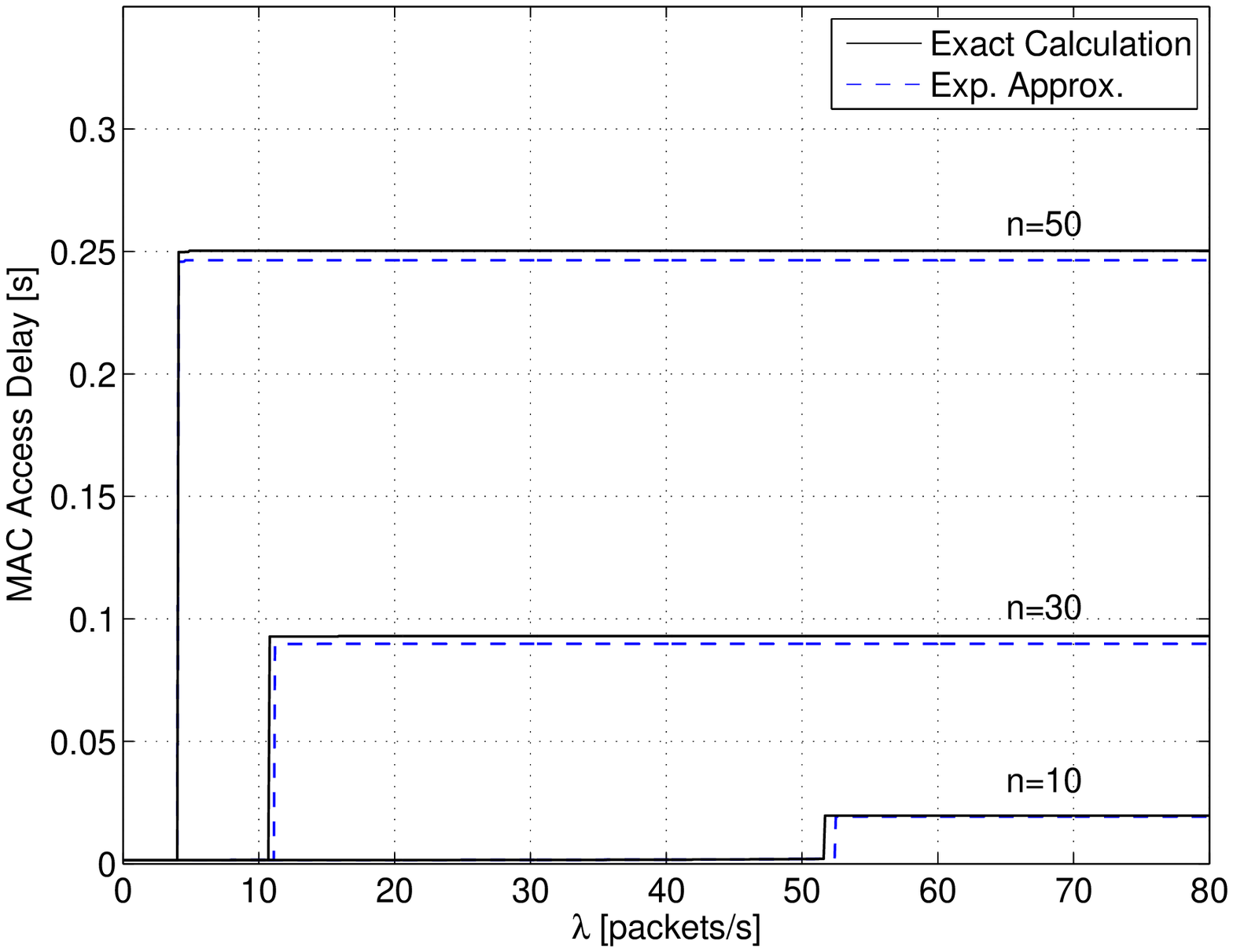}\label{fig:unsat_d_ca3ca2_exp}}\\
\subfigure[Throughput CA1/0]{\includegraphics[width=2.5in]{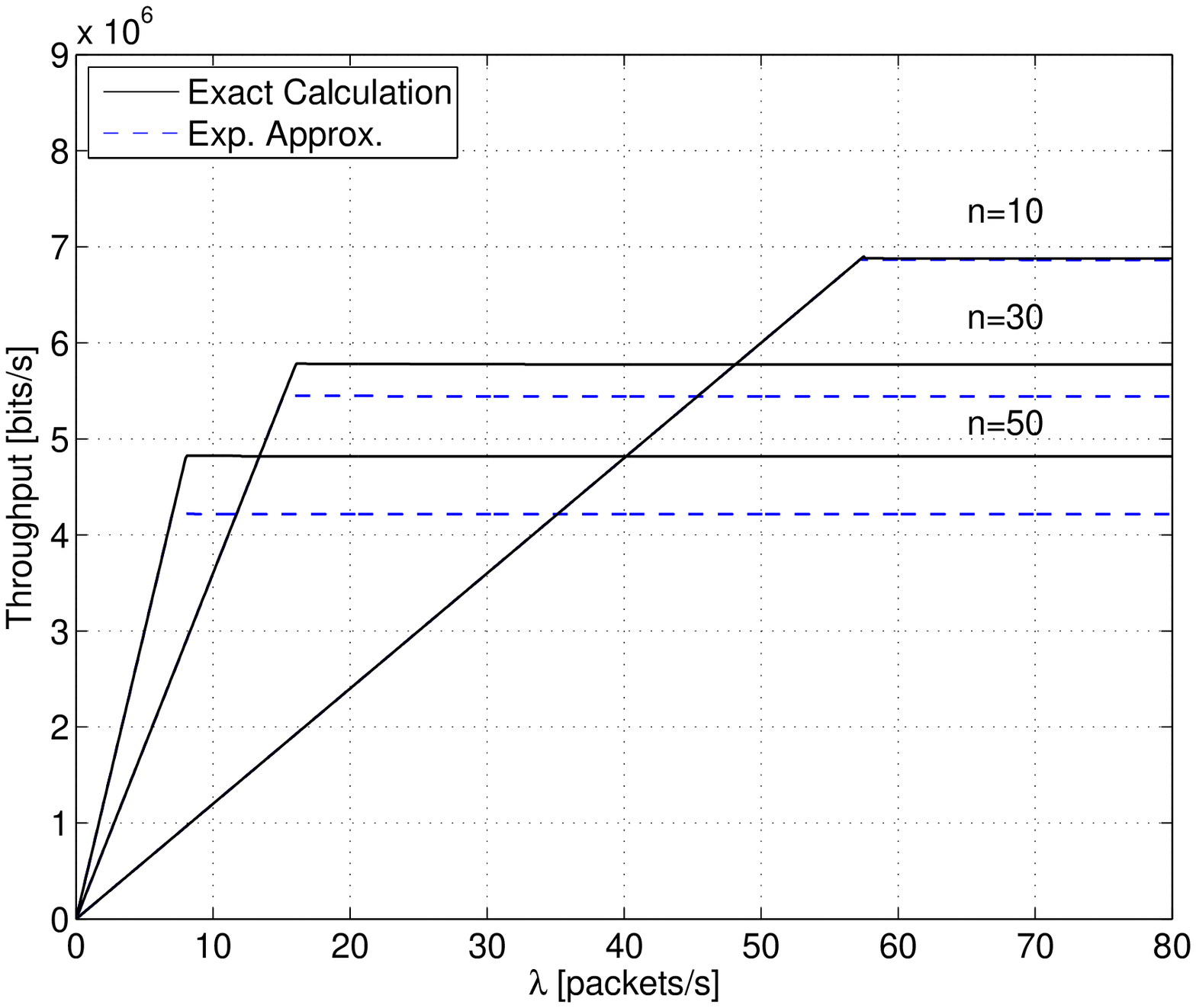}\label{fig:unsat_s_ca1ca0_exp}}
\subfigure[MAC access delay CA1/0]{\includegraphics[width=2.6in]{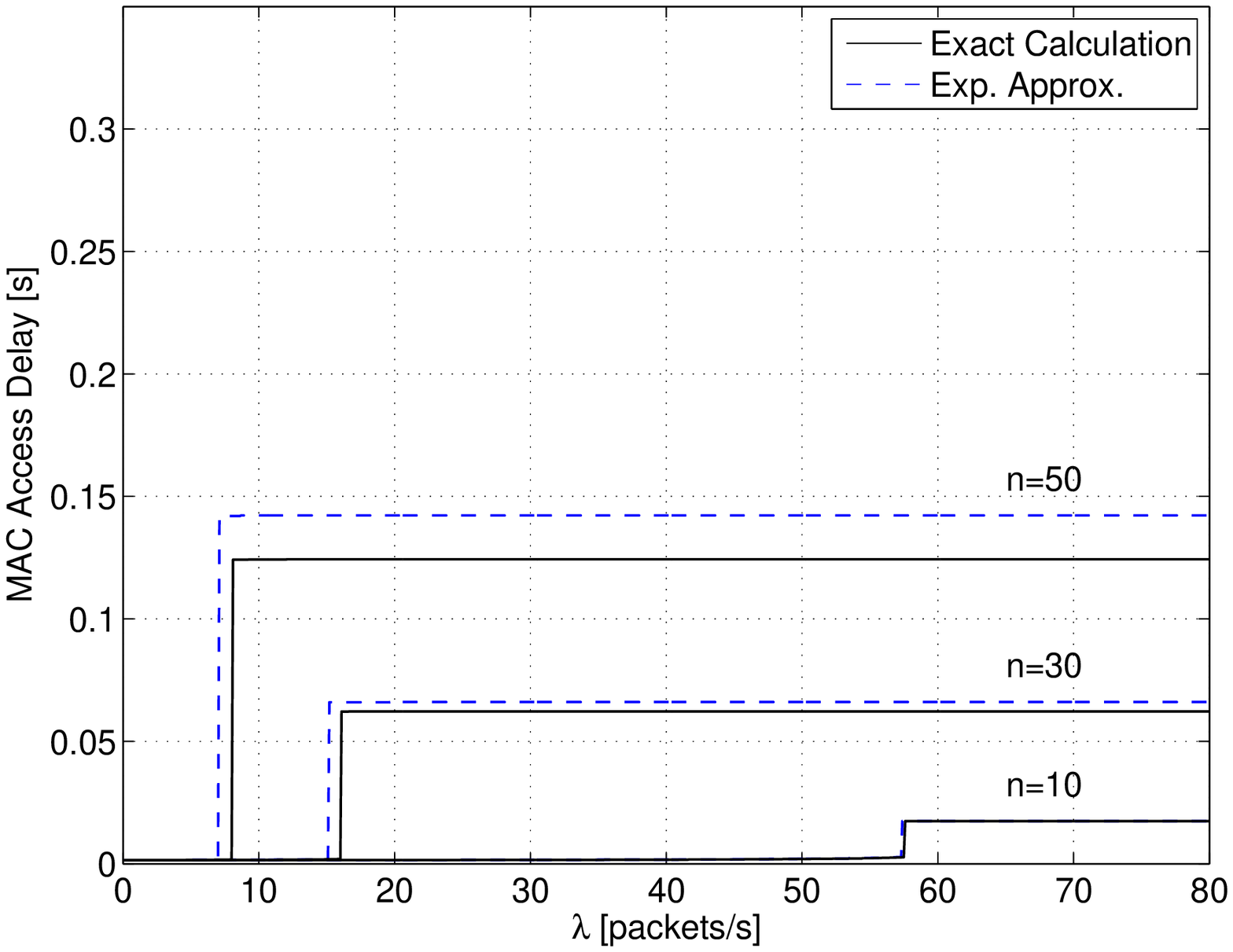}\label{fig:unsat_d_ca1ca0_exp}}
\caption{Performance results in unsaturated conditions. Comparison among the exact analysis and the exponential approximation to $p_{\rm defer}^{(i)}$.}
\label{fig:unsat_exp}
\end{figure*}

Finally, we compare the optional exponential approximation and the exact calculation (both starting with $I=0$ slots) in unsaturated conditions in Fig.~\ref{fig:unsat_exp}. When the system is saturated, we obtain the accuracy presented in the last subsection. Here it can be observed how the exponential approximation is able to predict the saturation point with a difference, for all cases evaluated, smaller than $\Delta\lambda=2$ packets/s.

\subsection{Effect of the Deferral Counter}

We now evaluate the effect of the the deferral counter on the difference between the two different solutions obtained from the analytic model right after the stability limit. These results allow us to quantify the potential misprediction errors. We consider \emph{i)} the results with no deferral (i.e., corresponding to a DCF-like access procedure \cite{IEEE80211-IEEESTD1999}) and \emph{ii)} increasing the backoff stage every time a packet from a neighbouring node is overheard (i.e., the proposal presented in \cite{campista2005improving} for improving the performance of Homeplug MAC). 

\subsubsection{No deferral}

\begin{figure*}[!tb]
\centering
\subfigure[Throughput CA3/2]{\includegraphics[width=2.6in]{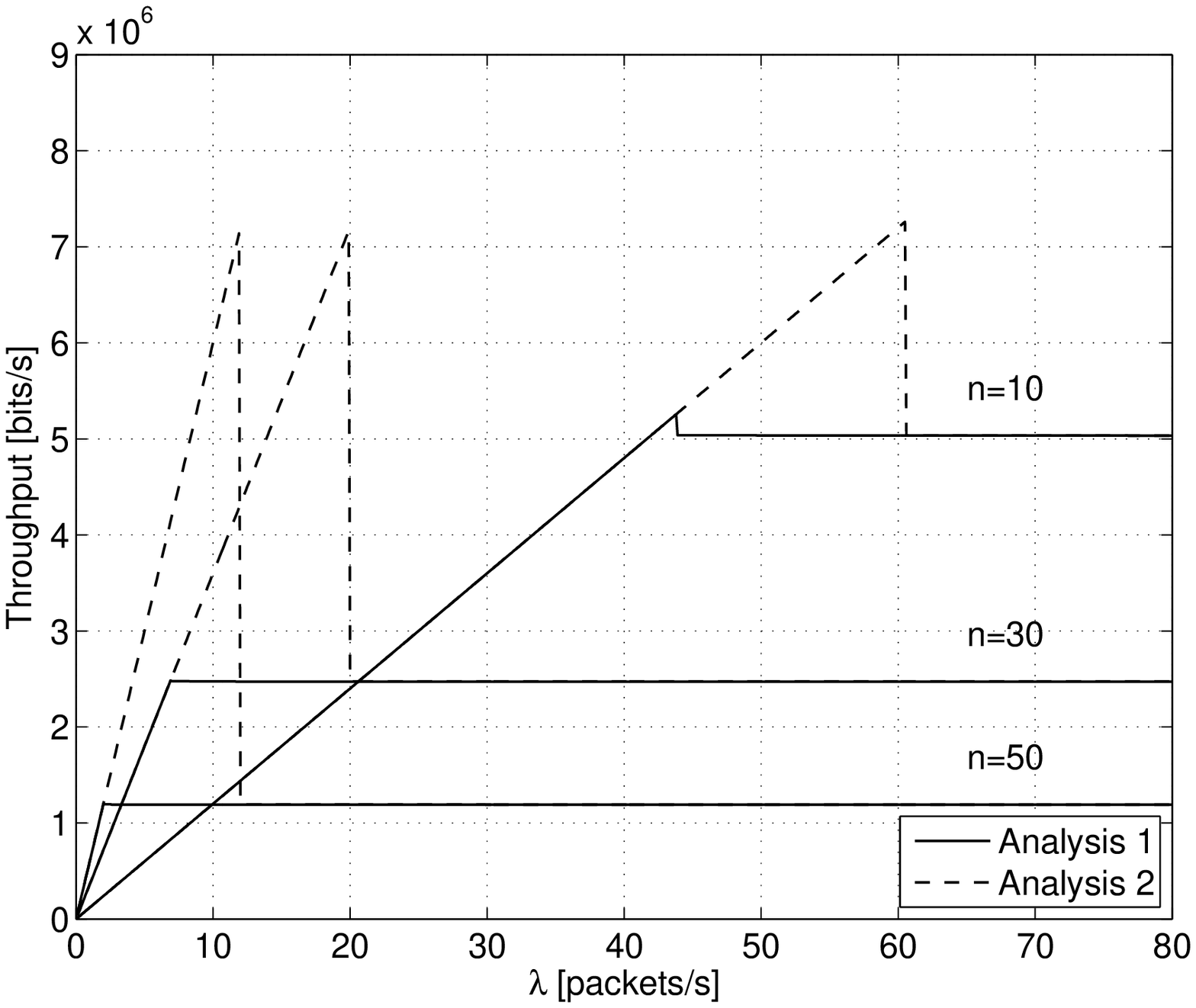}\label{fig:unsat_s_ca3ca2_no_deferral}}
\subfigure[Throughput CA1/0]{\includegraphics[width=2.6in]{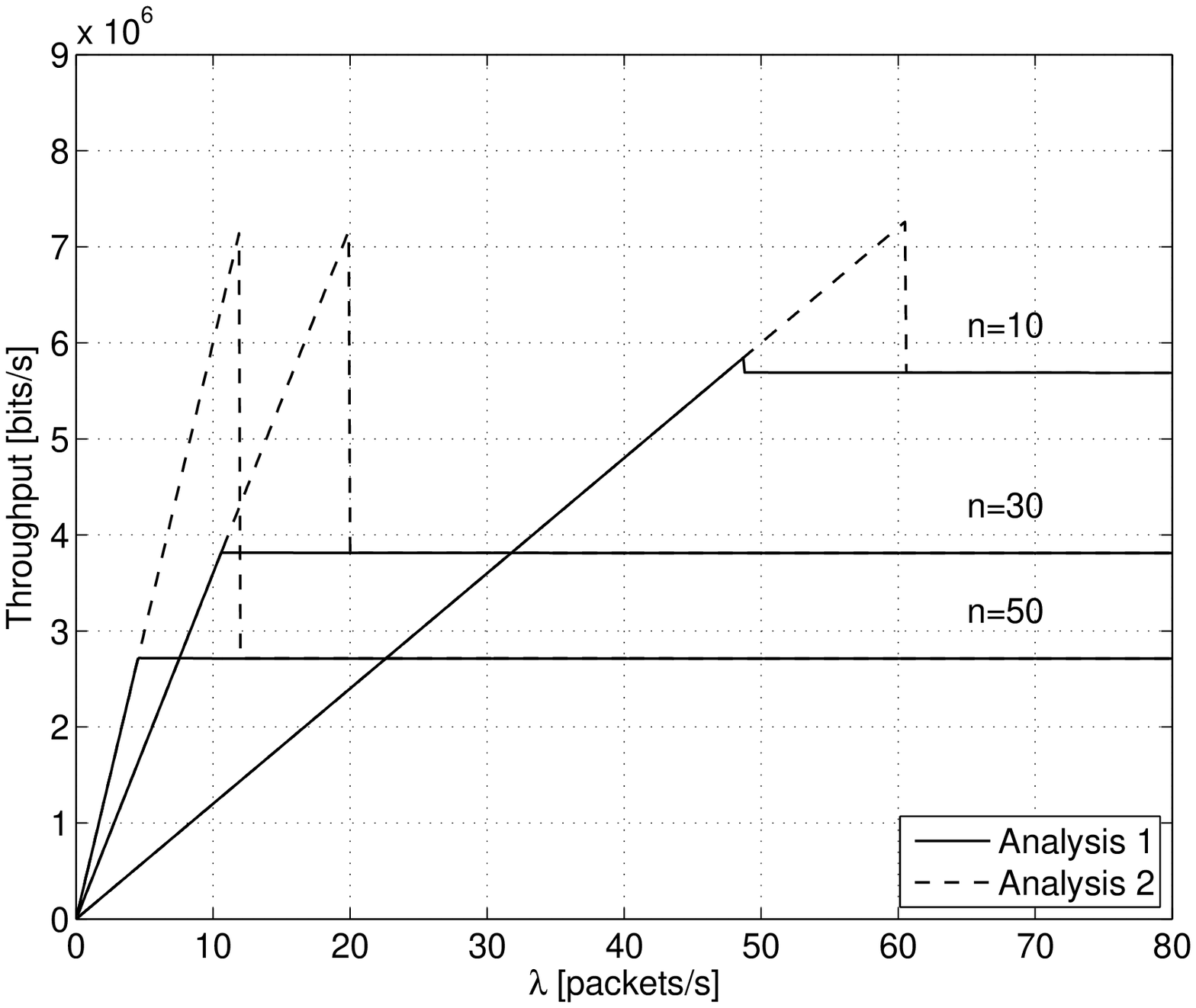}\label{fig:unsat_s_ca1ca0_no_deferral}}
\caption{Throughput in unsaturated conditions with no deferral ($M_i = \infty,~ \forall i$). Comparison among the two solutions derived from the exact analysis.}
\label{fig:unsat_no_deferral}
\end{figure*}

To obtain the results without the deferral counter (i.e., $M_i = \infty,~ \forall i$), we run the analytic model described in Section \ref{sec:model} with $p_{\rm defer}^{(i)} = 0, \forall i$ and $E[w_i] = (W_i+1)/2$. Note that the performance obtained will match that of DCF with the same parameters. However, for comparison purposes, we use the MAC parameters of Homeplug MAC shown in Table \ref{tbl:parameters}. We start the numerics for solving the model as in the last subsection: \emph{Analysis 1} corresponds to an initial value of $I = 0$ slots and \emph{Analysis 2} corresponds to the results starting with $I=1000$ slots. The two different solutions for throughput obtained from the analytic model are shown in Fig. \ref{fig:unsat_no_deferral}. Note that compared to the results depicted in Fig. \ref{fig:unsat}, the difference between the two solutions provided by the analytical model is higher in Fig. \ref{fig:unsat_no_deferral}. The saturation throughput obtained with no deferral is smaller compared to the one obtained using the deferral counter. Thus, with no deferral (Fig. \ref{fig:unsat_no_deferral}), the saturation throughput (\emph{Analysis 1}) and the throughput obtained under a lower contention assumption (\emph{Analysis 2}) show a bigger difference than the one found with deferral (Fig. \ref{fig:unsat}). Thus, under the same configuration, protocols such as the DCF have the potential to be affected by higher misprediction errors than Homeplug MAC, especially when small values of $W$ are used as 
in the case of CA3/2 categories. 

\subsubsection{Deferring always after overhearing }

\begin{figure*}[!tb]
\centering
\subfigure[Throughput CA3/2]{\includegraphics[width=2.6in]{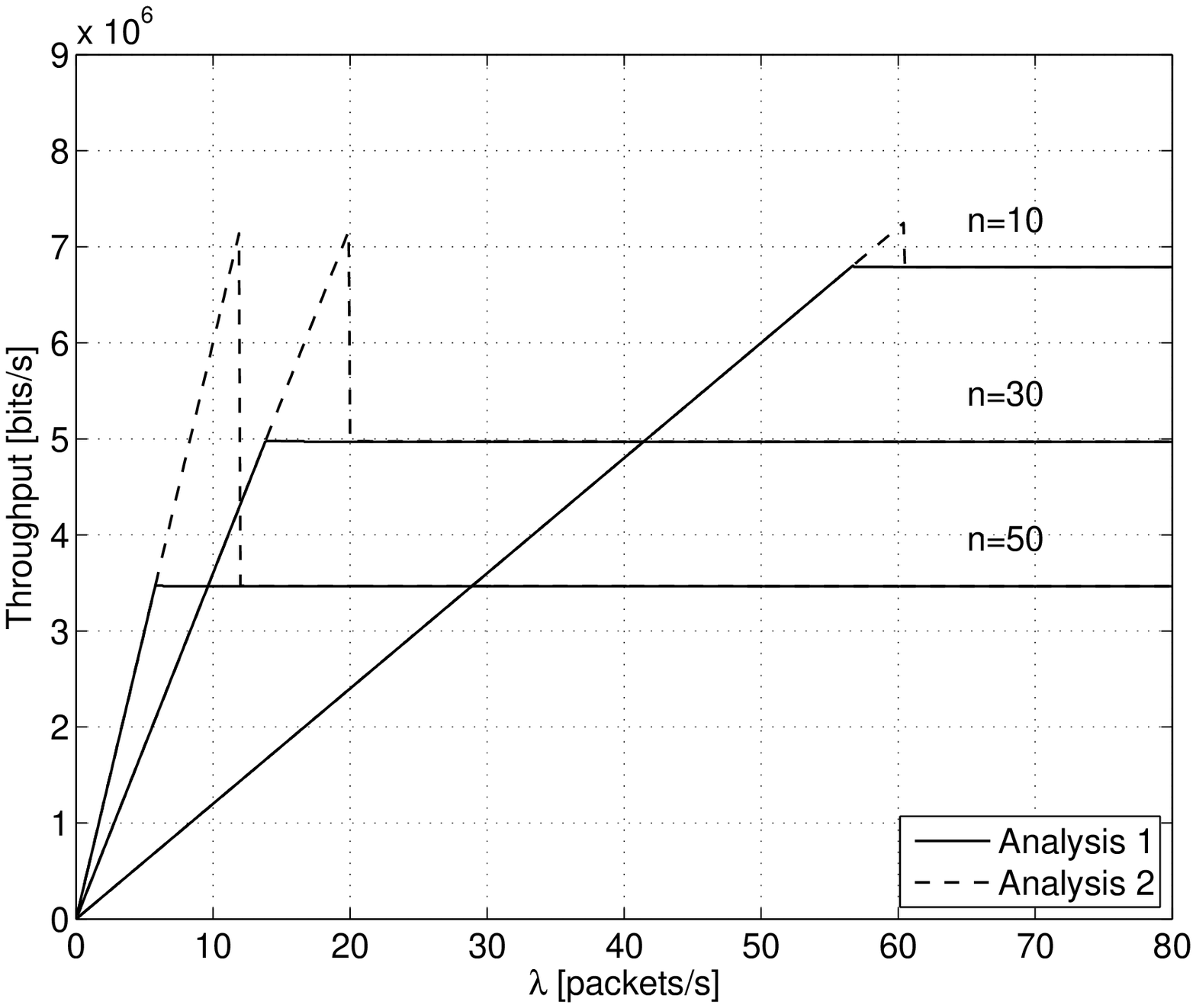}\label{fig:unsat_s_ca3ca2_always_defer}}
\subfigure[Throughput CA1/0]{\includegraphics[width=2.6in]{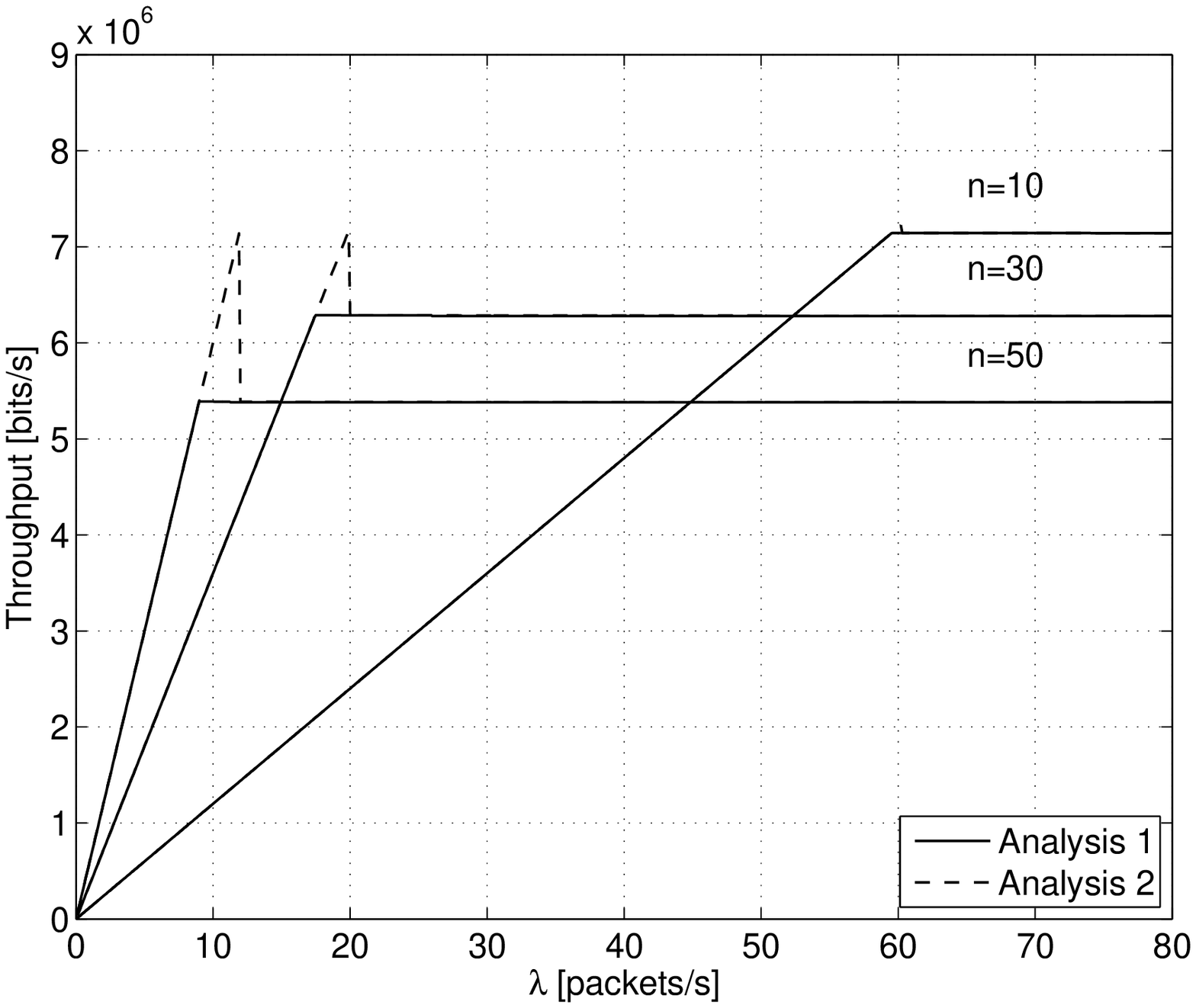}\label{fig:unsat_s_ca1ca0_always_defer}}
\caption{Throughput in unsaturated conditions deferring always after overhearing ($M_i = 0,~ \forall i$). Comparison among the two solutions derived from the exact analysis.}
\label{fig:unsat_always_defer}
\end{figure*}

In accordance with the conclusions drawn for no deferral, we expect that deferring always after overhearing will show a smaller difference among the two solutions as this setting increases the saturation throughput with higher contending nodes, especially when considering an increased number of backoff stages and larger contention windows. The improvement in the saturation throughput by setting $M_i=0,~ \forall i$ has already been studied for different configurations of $i$, $W$ and payload lengths (see Fig. 3 in \cite{campista2005improving}). However, in order to perform a direct comparison with the outcomes presented in previous sections, we show in Fig. \ref{fig:unsat_always_defer} the results using the same parameters considered throughout this work (i.e., those shown in Table \ref{tbl:access_categories} and \ref{tbl:parameters}) instead of referring to results in \cite{campista2005improving}. Moreover, this configuration allows us to evaluate the benefit of increasing the backoff stage after every packet overheard when the number of backoff stages are set to those recommended by the standard.

Observe that, under these particular settings, the saturation throughput compared to Fig. \ref{fig:unsat} is substantially improved but the potential misprediction errors are still considerable. 

\subsection{Reduction of Complexity}

Here, we evaluate the reduction in complexity obtained using our reformulated analysis (with precomputed $p_{\rm defer}^{(i)}$ and $E[w_i]$) and also our optional exponential approximation in comparison to the analysis presented in \cite{chung2006performance}. Moreover, we also compare the time needed to run the analytic models and $10000$~s-long simulations, as a crucial value of analysis is as a fast way to predict performance metrics. Since the analytic models are solved by means of iterative numerical methods, a regular complexity analysis based on the number of operations is not useful as the number of iterations needed may vary and cannot be predicted in advance. For this reason, we have obtained the elapsed times to run the analytic models (using \emph{tic} and \emph{toc} commands in Matlab R2011b) and simulations (using the \emph{time} command in Ubuntu). Tests have been performed in an Intel Core i5 at 3 GHz with 6 GB of RAM running a 12.04 32-bit Ubuntu operating system. Results of a single run considering $10$ and $50$ nodes using access categories CA1/0 and in saturated conditions are depicted in Table \ref{tbl:cpu_runtimes}. Obviously, these results may be affected by implementation issues. However, as can be observed, the improvement of our simplified analysis (both the exact analysis and the exponential approximation) is 2 orders of magnitude. Furthermore, the analysis in \cite{chung2006performance} has been shown to be of a similar computational expense to simulations, placing reasonable doubts on its efficiency.

\begin{table}
\centering
\begin{tabular}{ccccc} 
$n$ & Analysis in \cite{chung2006performance} & Exact Analysis & Exp. Approx. & Simulations\\ \hline
10 & $584.5$~s & $3.7$~s & $1.7$~s & $165.5$~s \\ \hline
50 & $420.0$~s & $4.2$~s & $3.5$~s & $866.2$~s \\ \hline
\end{tabular}
\caption{Elapsed Runtime Comparison for 10/50 nodes}\label{tbl:cpu_runtimes}
\end{table}

\section{Final Remarks}\label{sec:remarks}

In this work we have reformulated the analytic model of the Homeplug MAC procedure presented in \cite{chung2006performance} while maintaining its accuracy. To that end, we have applied a renewal reward approach to model the channel attempt rate instead of solving for the state probabilities associated with the Markov chain. Furthermore, following our approach, the most computationally expensive operations can be precomputed. We have considerably reduced the complexity of the analysis, resulting in a two orders-of-magnitude improvement in runtime. An optional exponential approximation has also been proposed and shown to be generally accurate for access categories CA3/2 and accurate under a small-conditional-collision-probability condition for access categories CA1/0.

Building upon previous literature on mean field analysis of random access MAC protocols and on stability analysis of systems of parallel queues with state-dependent service rates, we have demonstrated that the results right before the predicted saturation point in \cite{chung2006performance} correspond to a higher-throughput transitory phase of the network. We have also experimentally supported this finding and shown the extremely long duration of the transitory period. 

The reformulated analytic model has been validated in saturated and unsaturated conditions using simulations. Furthermore, we have found agreement (except right after the stability limit, as previously discussed) with the results presented in \cite{chung2006performance}, establishing the good accuracy of our proposal. In unsaturated conditions, we have shown the two solutions that the analytic model provides right after the stability limit and the long-term behaviour the network converges to by means of experimental evaluation. As expected, the long-term performance is in agreement with one of the solutions provided by the analytic model. 

In this article we have presented the first long-term performance evaluation of Homeplug MAC under infinite (or large enough to be considered infinite) buffer size. We have also highlighted the complexity of analysing a network right after the stability limit as analytic models that consider the decoupling approximation can provide two different solutions and simulations have to be run for a long time to allow the system to move to its long-term behaviour. Moreover, we have also provided techniques to obtain the long-term performance, both analytical and experimentally. Computing the maximum service rate the network can support allows us to identify the range of packet arrival rates for which a potential for two solutions exist. We have also suggested how to set the starting parameters of iterative numerical solvers to obtain both solutions. Experimentally, we have shown that the transitory phase can be drastically ameliorated if stations begin with a number of packets preloaded in their buffers. Indeed, 
this is a useful technique to identify MAC protocols which may be subject to a transitory phase via simulation. 

The potential occurrence of the extremely long transitory phase in generic random access protocols, as well as the characterisation of its duration, are important aspects to be analysed. They can provide further insights into the behaviour of networks based on random access protocols and serve as practical information to perform experimental evaluations. However, this study has been considered out of the scope of this work due to the complexity of analysing a system of coupled queues with state-dependent service rates, the reader is referred to \cite{cano2014longtrans} instead.

We have also evaluated the effect of the deferral counter on the magnitude of misprediction errors and shown that, under the same conditions, protocols such as the DCF have the potential to be affected by higher misprediction errors than the Homeplug MAC. On the other side, improvements of Homeplug MAC that increase the saturation throughput as the number of nodes increases show smaller misprediction errors.

\section*{Acknowledgment}

This work has been partially supported by the Science Foundation Ireland grant 08/SRC/I1403 and 07/SK/I1216a.

\bibliographystyle{elsarticle-num}
\bibliography{library}

\vspace{1cm}

\begin{wrapfigure}{l}{30mm}
    \includegraphics[width=30mm]{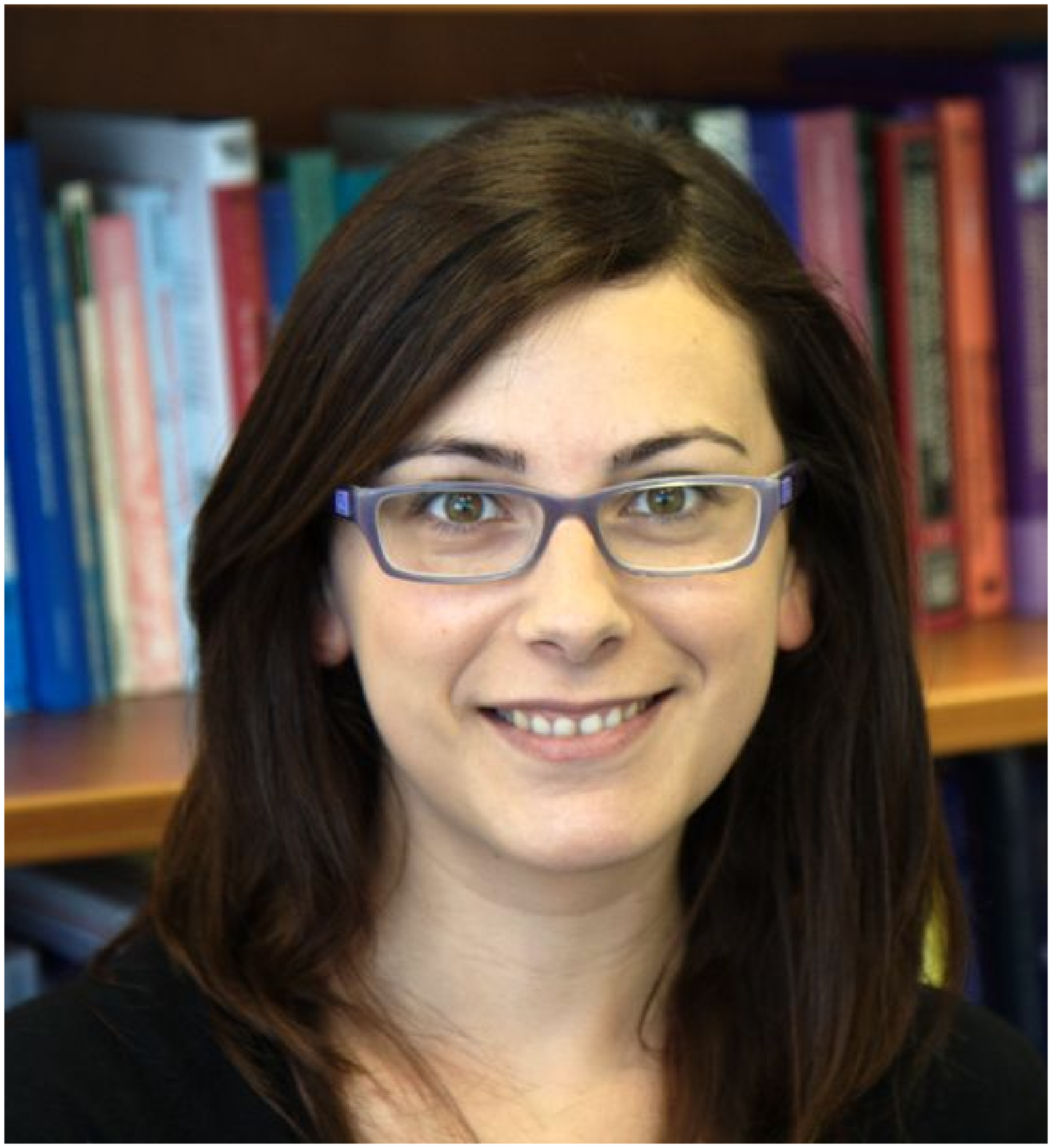}
\end{wrapfigure}
\textbf{Cristina Cano} obtained the Telecommunications Engineering Degree at the Universitat Politecnica de Catalunya (UPC) in February 2006. Then, she received her M.Sc. (2007) and Ph.D. (2011) on Information, Communication and Audiovisual Media Technologies from the Universitat Pompeu Fabra (UPF). Since July 2012, she has been working as a research fellow at the Hamilton Institute (NUIM) in wireless networks, sensor networks, power line communications and MAC layer design.\\

\begin{wrapfigure}{l}{30mm}
    \includegraphics[width=30mm]{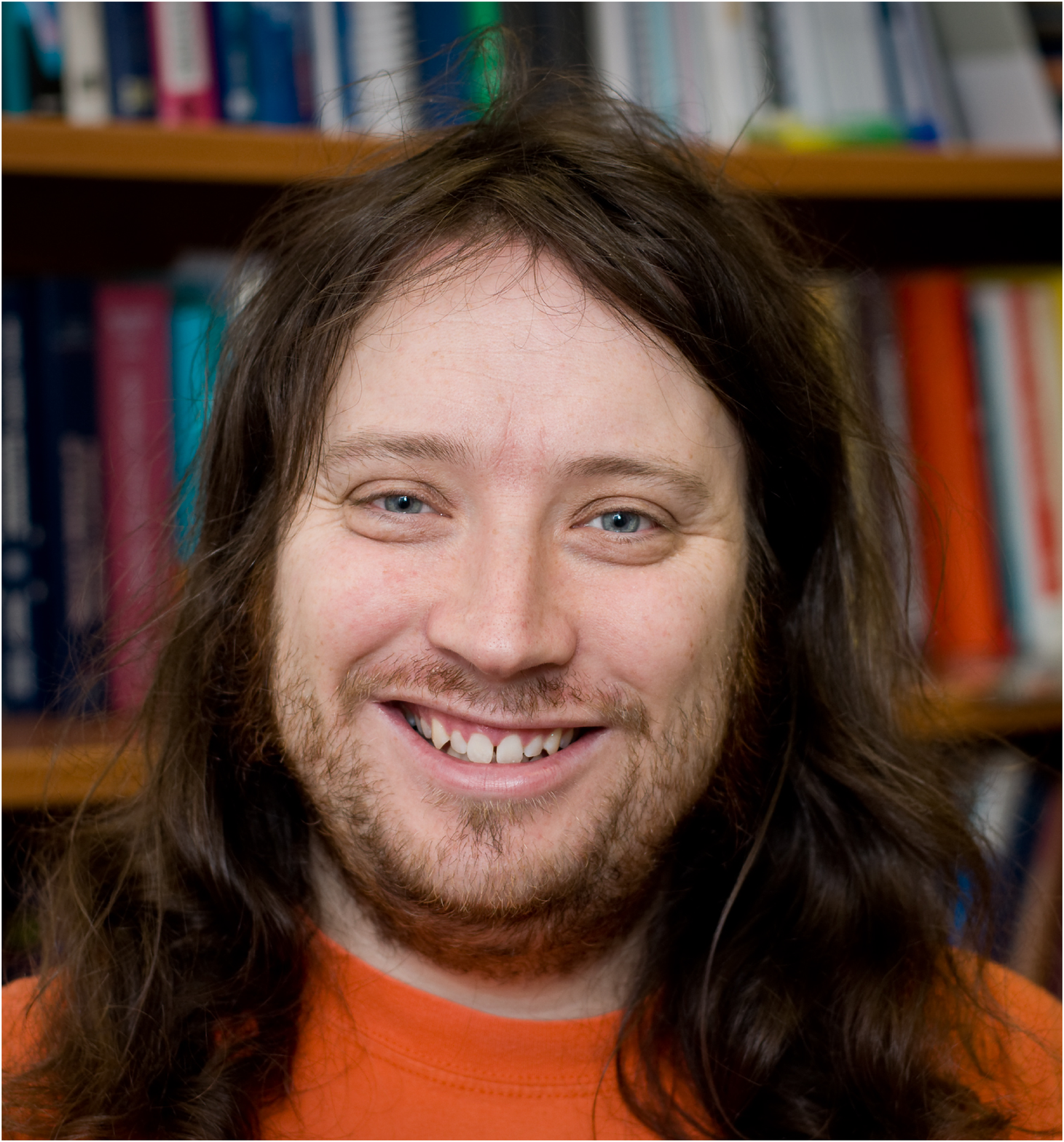}
\end{wrapfigure}
\textbf{David Malone} received B.A. (mod), M.Sc. and Ph.D. degrees in mathematics from Trinity College Dublin. During his time as a postgraduate, he became a member of the FreeBSD development team. He is currently a SFI Stokes Lecturer at the Hamilton Institute, NUI Maynooth. His interests include mathematics of networks, network measurement, IPv6 and systems administration. He is a coauthor of \emph{O'Reilly's IPv6 Network Administration}.

\end{document}